\documentclass[preprint]{emulateapj}

\def\oiii{\ifmmode [O {\sc iii}] \else [O {\sc iii}]\ \fi}
\def\Neii{\ifmmode [O {\sc ii}] \else [Ne {\sc ii}]\ \fi}
\def\Neiii{\ifmmode [O {\sc iii}] \else [Ne {\sc iii}]\ \fi}
\def\Nev{\ifmmode [O {\sc v}] \else [Ne {\sc v}]\ \fi}
\def\oiv{\ifmmode [O {\sc iv}] \else [O {\sc iv}]\ \fi}

\newcommand{\kms}{km~s$^{-1}$}

\begin{document}

\title{The different nature in Seyfert 2 galaxies with and
without hidden broad-line regions}

\author{Yu-Zhong Wu\altaffilmark{1,}\altaffilmark{2,}\altaffilmark{3},
En-Peng Zhang\altaffilmark{1,}\altaffilmark{3}, Yan-Chun
Liang\altaffilmark{1,}\altaffilmark{3}, Cheng-Min
Zhang\altaffilmark{1,}, and Yong-Heng
Zhao\altaffilmark{1,}\altaffilmark{3}}

\altaffiltext{1}{National Astronomical Observatories, Chinese
Academy of Sciences, 20A Datun Road, Beijing 100012, China.}
\altaffiltext{2}{Graduate University of Chinese Academy of Sciences,
19A Yuquan Road, Beijing 100049, China.} \altaffiltext{3}{Key
Laboratory of Optical Astronomy, National Astronomical
Observatories, Chinese Academy of Sciences, Beijing 100012, China.}
\email{yzhao@nao.cas.cn}

\shorttitle{The different dominant mechanisms in Sy2s}

\shortauthors{Wu et al}

\slugcomment{}

\begin{abstract}
We compile a large sample of 120 Seyfert 2 galaxies (Sy2s) which
contains 49 hidden broad-line region (HBLR) Sy2s and 71 non-HBLR
Sy2s. From the difference in the power sources between two groups,
we test if HBLR Sy2s are dominated by active galactic nuclei (AGNs),
and if non-HBLR Sy2s are dominated by starbursts.
 We show that: (1) HBLR Sy2s have larger accretion
rates than non-HBLR Sy2s; (2) HBLR Sy2s have larger \Nev $\lambda
14.32$/\Neii $\lambda 12.81$ and \oiv $\lambda 25.89$/\Neii $\lambda
12.81$ line ratios than non-HBLR Sy2s; (3) HBLR Sy2s have smaller
$IRAS$ $f_{60}/f_{25}$ flux ratio which shows the relative strength
of the host galaxy and nuclear emission than non-HBLR Sy2s. So we
suggest that HBLR Sy2s and non-HBLR Sy2s are AGN-dominated and
starburst-dominated, respectively. In addition, non-HBLR Sy2s can be
classified into the luminous ($L_{\rm [O~III]}>10^{41}
\rm~ergs~s^{-1}$) and less luminous ($L_{\rm [O~III]}<10^{41}
\rm~ergs~s^{-1}$) samples, when considering only their obscuration.
We suggest that: (1) the invisibility of polarized broad lines
(PBLs) in the luminous non-HBLR Sy2s depends on the obscuration; (2)
the invisibility of PBLs in the less luminous non-HBLR Sy2s depends
on the very low Eddington ratio rather than the obscuration.

\end{abstract}
\keywords{galaxies: active --- galaxies: Seyfert --- galaxies: statistics}

\section{INTRODUCTION}

According to the unified model of active galactic nuclei (AGNs;
Antonucci 1993), type 1 AGNs are seen face-on and have both narrow
and broad emission lines; type 2 AGNs are seen edge-on and have only
narrow emission lines, which are commonly believed to be
intrinsically the same as type 1 AGNs. With the discoveries of both
polarized broad lines (PBLs) in NGC 1068 (Antonucci \& Miller 1985)
and some hidden broad-line regions (HBLRs) in other Seyfert 2
galaxies (Sy2s; Miller \& Goodrich 1990; Tran et al. 1992; Young et
al. 1996; Heisler et al. 1997; Kay \& Moran 1998), Seyfert 2
galaxies are classified into HBLR and non-HBLR Sy2s. About $50\%$ of
the total currently known Seyfert 2 galaxies show the presence of
HBLRs in their polarized optical spectra, while the remaining half
do not (Tran 2001, 2003; Nicastro et al. 2003; Haas et al. 2007).

It seems to have an indication that the activities of the two kinds
of objects may be powered by different mechanisms. Based on the
results of a spectropolarimetric survey of the CfA and $12~\mu $m
samples of Sy2s, Tran (2001)
 proposed the existence of a population of galactic
 nuclei whose activity is powered by starburst rather than accretion
 onto a supermassive black hole (SMBH) and in which, therefore, the BLRs
 simply do not exist (Nicastro et al. 2003). With respect to the radio,
  far-infrared, and near-infrared emissions of the two groups, Yu \& Hwang (2005) found
that an HBLR Sy2 is similar to a Sy1, suggesting that this type of
object does harbor a central AGN; on the other hand, the non-HBLR
Sy2 is more like a starburst galaxy.

Considerable efforts have been devoted in the past decade to
understanding the HBLR and non-HBLR Sy2s. The absence of PBLs could
be attributed to the edge-on line of sight and hidden of electron
scattering region (Heisler et al. 1997; Wang $\&$ Zhang 2007). Many
evidences showed that the presence or absence of HBLRs in Seyfert 2
galaxies depends on the AGN luminosity, with the HBLR sources
having, on average, larger luminosities (Lumsden \& Alexander 2001;
Gu \& Huang 2002; Martocchia \& Matt 2002; Tran 2001, 2003; Nicastro
et al. 2003). Examining the sample extracted from the
spectropolarimetric survey of Tran (2001, 2003), Nicastro et al.
(2003) found that all HBLR sources have accretion rates larger than
the threshold value of $\dot{m}\simeq10^{-3}$(in Eddington units),
while non-HBLR sources lie at $\dot{m}\le {\dot{m}}_{\rm thres}$.
Collecting a sample of 90 Sy2s with radio, infrared, optical, and
X-ray (2-10 keV) data, Gu \& Huang (2002) indicated that the
majority of non-HBLR Sy2s have less powerful AGN activity, which is
likely caused by a low accretion rate. Based on the observed upper
limit of emission line width of 25,000 \kms, Laor (2003) also
proposed a model to describe the existence of BLRs in AGNs.

Seyfert 2 galaxies have large columns of circumnuclear obscuring
material that prevents the direct view of the nucleus. X-ray
observations are useful for providing an indication of the level of
obscuration by the torus. One usually uses the column density of
neutral hydrogen ($N_{\rm H}$) to show the obscuration. In the local
universe, about half of Sy2s are found to be Compton-thick sources
with $N_{\rm H}> 10^{24}$ cm$^{-2}$ (Maiolino et al. 1998; Bassani
et al. 1999; Risaliti et al. 1999). However, some Sy2s do not show
HBLRs in spectropolarimetric observations and have column densities
lower than $10^{22} \rm cm^{-2}$ in the X-ray observations (Panessa
\& Bassini 2002), which indeed challenge the unified model (Bian \&
Gu 2007).

About the nature of the power sources in HBLR and non-HBLR Sy2s,
there are still some controversies. Moreover, the reason that Sy2s
with column densities lower than $10^{22} \rm cm^{-2}$ do not show
HBLRs is still unclear. In this paper, therefore we firstly devote
to distinguish between HBLR and non-HBLR Sy2s in dominant mechanisms
(AGNs or starbursts); then we investigate and discuss physical
reasons of the absence of PBLs in non-HBLR Sy2s. We assume $H_0$ =75
km $\rm s^{-1}$ Mpc$^{-1}$, $\Omega _{\rm M}=0.3$, and $\Omega _
\Lambda=0.7$ throughout the paper.

\section{THE SAMPLE and DATA}
We collect multi-wavelength data for the large sample of 120 Sy2s
which consists of 71 non-HBLR Sy2s and 49 HBLR Sy2s listed in Tables
1 and 2, respectively, including radio, far-infrared, infrared,
optical, and X-ray (2-10 keV) bands. The sample selection is mainly
from Gu \& Huang (2002), Tran (2003), Wang \& Zhang (2007), and Shu
et al. (2007). According to the Sy2 classification of Tran (2003),
we classify the two objects, NGC 5347 and NGC 5929, as the non-HBLR
Sy2 sample. Except the 18 objects in Table 5 of Wang \& Zhang
(2007), all other objects of our sample have their
spectropolarimetric observations which are described in Appendix in
detail. With regard to the 18 objects, Wang \& Zhang (2007) took the
two criteria to classify unabsorbed Seyfert 2 galaxies into non-HBLR
Sy2s and HBLR Sy2s (see section 2 of Wang \& Zhang 2007; 14 non-HBLR
Sy2s and 4 HBLR Sy2s).

\begin{table*}
\caption{\small The non-HBLR Sy2 sample}
\begin{small}
\setlength{\tabcolsep}{0.5pt}
\renewcommand{\arraystretch}{0.5}

\begin{tabular}{lccccccccccccccl} \hline \hline

Name & $z$ & $M_{\rm BH}$ & $f_{25}$ & $f_{60}$ & $f_{100}$ & \Neii
& \Nev & \oiv & $ L_{\rm [O~ III]}$
& $ L_{\rm 1.49 GHz}$ & $F_{\rm HX}$ & $\dot{M}$ &$ N_{\rm H}$& EW &Reference \\
      (1) & (2) &(3) & (4) & (5) & (6)&(7)& (8) & (9)& (10)& (11)&(12) &(13)& (14) &(15)&(16)\\
\hline

ESO 428-G014& 0.006 & 7.34 & 1.77 & 4.40 & 6.05 &  ...  &  82.9   &  ...   & 42.23 &28.542& 3.80 & 1.04 &$>$25.00&1600&3,3,1,3,5,4,3,4\\
F00198-7926 & 0.073 & ...  & 1.15 & 3.10 & 2.87 & 6.19  & 12.27   & 33.03  & 42.67 & ...  &$<$1.0& 2.86 &$>$24.00&....&2,19,4,4,4\\
F01428-0404 & 0.018 & 7.31&$<$0.34& 0.66 & 1.71 & ...   &   ...   & ...    & ...&29.143$^a$&...  &  ... & 21.51&....&7,1,1,7\\
F03362-1642 & 0.037 & ...  & 0.50 & 1.06 &2.01  & ...   &  ...    & ...    & 41.62 &29.370& ...  & 0.26 & ...  &....&3,3,5\\
F04103-2838 & 0.117 & ...  & 0.54 & 1.82 & 1.71 & ...   &  ...    &   ...  & ...   &30.539& 0.38 & ...  & ...  &....&1,5,15\\
F04210+0401 & 0.045 & 7.34 & 0.25 & 0.60 &$<$2.54&  ...  &  ...    &   ...  & 42.42&30.295& ...  & 1.61 & ...  &....&3,3,3,5\\
F04229-2528 & 0.044 & ...  & 0.26 & 0.98 & 1.25 &  ...  &  ...    &   ...  & 41.99 &29.547& ...  & 0.60 & ...  &....&3,3,5\\
F04259-0440 & 0.016 & ...  & 1.41 & 4.13 & 3.30 & ...   &  ...    &   ...  & 41.85 & ...  & ...  & 0.43 & ...  &....&1,4\\
F08277-0242 & 0.041 & ...  & 0.43 & 1.47 & 1.82 & ...   &  ...    &   ...  & 41.76 &30.028& ...  & 0.35 & ...  &....&3,3,5\\
F10340+0609 & 0.012 & ... &$<$0.25&0.39&$<$1.12 & ...   &  ...    &   ...  & ...   & ...  & 7.8& ...  & ...  &....&3,1\\
F13452-4155 & 0.039 & 6.52 & 0.81 & 1.84 & 1.34 & ...   &  ...    & ...    & 42.19 & ...  & ...  & 0.95 & ...  &... &3,3,3\\
F19254-7245 & 0.0617& ...  & 1.35 & 5.24 & 8.03 & 31.48 &  2.77   & 6.35   & 43.06 & ...  & 2.3  & 7.03 &$>$24 &2000&2,18,4,4,4,4\\
F20210+1121 & 0.056 & ...  & 1.40 & 3.39 & 2.68 & ...   & ...     &   ...  & 43.31 &30.485& 3.0  & 12.51&$>$25.00&1650&1,4,5,11,4,4\\
F23128-5919 & 0.045 & ...  & 1.59 & 10.8 & 11   & 27.29 &  2.56   & 18.16  & 41.68 & ...  & 1.3  & 0.29 &22.681&....&3,18,4,4,4\\
IC 5298     & 0.027 & ...  & 1.80 & 9.76 & 11.1 & ...   &  ...    &  ...   & 42.17 &29.715& ...  & 0.91 & ...  &....&3,3,5\\
Mrk 334     & 0.022 & 6.52 & 1.05 & 4.35 & 4.32 & 30.0  & 13.0    &   15.0 & 42.34 &29.434&$<$130& 1.34 &20.643&....&9,1,17,4,5,4,4\\
Mrk 573     & 0.017 & 6.04 & 0.85 & 1.24 & 1.43 & ...   &  ...    &  ...   & 42.39 &29.133& 1.2 & 1.50&$>$24.00&2800&3,1,3,5,4,3,4\\
Mrk 938     & 0.02  & 7.0  & 2.51 & 16.84& 17.61& 64.0  &  ...    &  ...   & 42.69 &29.705& 2.3 & 3.00 &$>$24.00&$<$321&3,3,19,3,5,4,4,4\\
Mrk 1066    & 0.012 & 7.5  & 2.26 & 11.0 & 12.2 &  ...  & 17.8   & ...     & 42.27 &29.467& 2.3 & 1.14&$>$24.0&1120&3,3,1,3,5,4,3,4\\
Mrk 1361    & 0.023 & ...  & 0.84 & 3.28 & 3.73 & ...   & ...     & ...    & 42.33 &29.297& ...  & 1.31 & ...  &....&3,3,5\\
NGC676      & 0.005 & 8.27 &$<$0.062& 0.27 & 0.80 & ...   &  ...    & ...    & 39.21 & ...&0.112&$<$0.001&$\le$21.00&....&7,1,7,16,7\\
NGC1058     & 0.002 & 6.03 & 0.17 & 2.65 & 8.74 & ...   &  ...    &  ...   & 38.06&26.928$^a$&0.024&0.0001&$\le$21.78&...&7,8,7,1,16,7\\
NGC1143     & 0.029 & ...&$<$0.10&$<$1.10&$<$1.5&   15.0&  ...    &  ...   & 41.97 & ...  & ...  & 0.57 & ...  &....&13,1,4\\
NGC1144     & 0.029 & 5.84 & 0.62 & 5.35 & 11.6 &  ...  & ...     &  ...   & 41.81 &30.391&$<$120& 0.40 & 22.00&....&3,3,3,5,4,3\\
NGC1241     & 0.014&$<$7.34& 0.60 & 4.37 & 10.74&   9.0 &  ...    & 2.0    & 41.74 & ...  & ...  & 0.34 & ...  &....&3,3,17,3,\\
NGC1320     & 0.009 & 6.36 & 1.32 & 2.21 & 2.82 & 9.0   & 8.0     &  32.0  & 41.08 &27.974&$<$82.0& 0.07 & ...  &....&3,3,17,3,5,4\\
NGC1358     & 0.013 & 6.29&$<$0.12& 0.38 & 0.93 &  ...  &  ...    &  ...   & 41.36 &28.820&8.6   & 0.14 & 23.60&....&3,3,3,5,4,4\\
NGC1386     & 0.003 & 6.92 & 1.46 & 6.01 & 9.67 &  28.0 & 45.4    & ...    & 41.09 &27.637& 2.0 & 0.08 & 25.00&7600&3,3,1,3,5,11,3,2\\
NGC1667     & 0.015 & 6.68 & 0.67 & 6.29 & 15.83&  26.0 & ...     &   12.0 & 42.03 &29.506& 0.26& 0.66 &$>$24.00&$<$3000&3,3,17,3,5,11,4,2\\
NGC1685     & 0.015 & ...  & 0.22 & 0.98 & 1.53 & ...   &  ...    &  ...   & 42.67 &28.753&$<$20.0& 2.86 & ...  &....&3,3,5,4\\
NGC2685     & 0.003 & 6.97&$<$0.11& 0.37 & 1.66 & ...   &  ...    & ...    & 38.92&26.743$^a$&2.70&0.0005&$\le$21.48&....&7,8,7,1,1,7\\
NGC3031     & 0.001 & 7.83 & 5.42 & 44.73&174.02& ...   &   0.6   &  ...   & 38.81 &27.504$^a$&150&0.0004&$\le$21.00& 170$^d$&7,8,1,7,1,11,7,11\\
NGC3079     & 0.004 & ...  & 3.65 & 50.95& 105.2& 148.0 &   0.7   &   36.0 & 40.48 &29.359& 5.3  & 0.02 & 25.00&1480&3,$19^b$,3,5,11,4,4\\
NGC3147     & 0.009 & 8.64 & 1.08 & 8.40 & 29.96& ...   & ...     &  ...   & 40.19 &29.350$^a$&13.0& 0.01&$\le$20.46& 485&7,8,7,1,11,7,11\\
NGC3281     & 0.012 & 6.41 & 2.63 & 6.73 & 7.89 & ...   & ...     & ...    & 41.30 &29.248& 40.0& 0.12 & 24.20&1180&3,14,3,5,11,4,4\\
NGC3362     & 0.028 & 7.20 & 0.35 & 2.13 & 3.16 &  ...  & ...     &  ...   & 41.37 &29.376&$<$126.0& 0.14 & ...  &....&3,3,3,5,4\\
NGC3393     & 0.013 & ...  & 0.75 & 2.25 & 3.87 & ...   &   42.4  &  ...   & 42.10 &29.643$^a$& 4.0& 0.77&$<$23.9&3500&3,1,3,1,11,3,11\\
NGC3486     & 0.002 & 6.17 & 0.32 & 6.24 & 15.87&  ...  &  ...    & ...    & 38.25 &27.827$^a$&0.85&0.0001&$\le$21.48&....&7,8,7,1,16,7\\
NGC3660     & 0.012 & 7.33 & 0.64 & 2.03 & 4.47 &  6.51 &    0.98 &  3.61  & 40.76 &28.513&22.0 &0.035  & 20.26&....&7,2,18,4,5,4,4\\
NGC3941     & 0.003 & 7.39 & ...  & ...  & ...  &  ...  &  ...    &  ...   & 38.80 & ...  &0.419&0.0004&$\le$21.00&....&7,7,16,7\\
NGC3982     & 0.004 & 6.15 & 0.97 & 7.21 & 16.78&   16.0&  ...    & 2.0    & 40.33 &28.226& 5.7 & 0.01 &$>$24.2&6310&3,3,17,3,5,4,3,4\\
NGC4117     & 0.003 & 6.03 &  ... & ...  &  ... & ...   &  ...    &  ...   & ...   &27.020&$<$232.0& ...  & ...  &....&3,5,4\\
NGC4472     & 0.003&8.67&$<$0.21&$<$0.19&$<$0.48& ...   &  ...    &  ...   & 37.62 &28.850$^a$&2.15&0.00003& 21.48&....&7,8,7,1,16,7\\
NGC4501     & 0.008 & 7.86 & 3.02 & 19.93&63.64 &  7.02 &   1.5   &   4.22 & 39.89 &29.468& 1.1 &0.005&$<$21.30&....&9,2,19,4,5,4,4\\
NGC4565     & 0.004 & 7.56 & 1.7  & 9.83 & 47.23&  ...  &  ...    &  ...   & 39.36 &28.818$^a$&2.07&0.0014& 20.11&....&7,8,7,1,16,7\\
NGC4579     & 0.005 & 7.74 & 0.72 & 6.70 & 18.92&   11.0&   0.6   &  ...   & 39.58 &28.871$^a$&44.0&0.0023& 20.39& 240&7,8,19$^c$,7,1,11,7,11\\
NGC4594     & 0.004 & 8.52 & 0.50 & 4.26 & 22.86& ...   &    0.3  &   ...  & 39.26 & ...  &19.0 & 0.001& 21.23&....&7,8,1,7,11,7\\
NGC4698     & 0.003 & 7.43 &$<$0.154&0.258&1.864&  ...  &  ...    &   ...  & 38.64 & ...  & 0.48 &0.0003& 20.91&$<$425&7,1,7,16,7,12\\
NGC4941     & 0.004 & 6.34 & 0.46 & 1.87 & 4.79 &  ...  &    9.0  &   19.0 & 41.18 &27.629& 7.0  & 0.09 & 23.65&1600&3,3,17,3,5,11,3,11\\
NGC5033     & 0.003 & 7.48 & 1.15 & 13.8 & 43.9 &  13.3 &   0.4   &   5.1  & 39.47 &28.992$^a$& 55.0 & 0.002& 20.01& 290&7,10,20,7,1,11,7,11\\
NGC5128     & 0.002 & 8.30 & 28.2 & 213.0&412.0 &  203.0&   22.0  &  124.0 & 38.82 & ...  & 850.0&0.0004&$>$23.0&114&21,3,19,3,11,3,11\\
NGC5135     & 0.014 & 5.79 & 2.39 & 16.6 &31.18 &  ...  &   25.2  &  ...   & 42.21 &29.824& 2.0  & 0.99&$>$24.0&$<$11700&3,3,1,3,5,11,3,2\\
NGC5194     &0.00154& ...  & 17.47& 108.7&292.08&   17.0&   0.74  &   7.9  & 40.03 &28.441& 11.0 & 0.007&24.748& 986&8,19,4,5,11,4,4\\
NGC5256     & 0.028 & 6.92 & 1.07 & 7.25 & 10.11&   76.0&   2.1   &   61.0 & 41.89 &30.283& 5.6  & 0.48&$>$25.0& 575&9,14,$17^b$,4,5,4,4,4\\
NGC5283     & 0.01  & 7.14 &0.089 & 0.132& 0.751& ...   &   ...   &  ...   & 40.88 &28.355& 14.6 & 0.05 & 23.18&$<$220&3,1,3,5,4,3,4\\
NGC5347     & 0.008 & 7.3  & 0.96 & 1.42 & 2.64 & 3.0   &   ...   &  4.0   & 41.22 &27.852& 2.2  & 0.10 &$>$24.00&1300&3,3,17,3,5,4,4,4\\
NGC5643     & 0.004 & 6.45 & 3.65 & 19.5 & 38.2 &  46.4 &  24.6   & 118.3  & 41.37 &28.944& 13.0 & 0.14 & 23.85& 500&3,3,20,3,1,11,3,4\\
NGC5695     & 0.014 & 7.15 & 0.13 & 0.57 & 1.79 &  ...  &  ...    &  ...   & 40.55 &28.397&$<$1.0& 0.02 & ...  &....&3,3,3,5,4\\
NGC5728     & 0.009 & 6.95 & 0.88 & 8.16 & 14.7 &   ... &  ...    & ...    & 41.09 &29.047& 13.3 & 0.08 & 23.89&1100&3,3,3,5,4,4,4\\
NGC5929     & 0.008 & 7.25 & 1.67 & 9.52 & 13.84&  21.0 &  4.0    &  7.0   & 41.40 &29.220& 1.35 & 0.15 & 22.63&....&9,2,17,4,5,4,4\\
NGC6251     & 0.023 & 8.8  & 0.07 & 0.19 & 0.60 &  ...  &  ...    &  ...   & 41.77 &31.582$^a$& 13.0 & 0.36 & 21.88& 392$^d$&7,1,7,1,11,7,11\\
NGC6300     & 0.004 & 6.29 & 2.27 & 14.7 & 36.0 &  11.5 &  12.5   &  29.5  & 41.08 & ...  & 216.0& 0.07 & 23.34& 148&3,3,20,3,4,3,4\\
NGC6890     & 0.008 & 6.48 & 0.65 & 3.85 & 8.16 &  11.32&    5.77 &  10.1  & 41.04 & ...  & 0.80 & 0.07 & ...  &....&3,3,19,3,1\\
NGC7130     & 0.016 & ...  & 2.16 & 16.71& 25.89&  71.0 &  11.0   &   43.0 & 42.55 &29.977& 1.60 & 2.17 &$>$24.00&1800&14,17,4,5,4,4,4\\
NGC7172     & 0.009 & 7.67 & 0.95 & 5.74 & 12.43&  ...  &  ...    &  50.0  & 40.84 &28.731& 130.0& 0.04 & 22.94& 121&9,3,17,3,5,11,3,2\\
NGC7496     & 0.005 & ...  & 1.93 & 10.14& 16.57& 48.08 &   1.8   &   2.4  & 40.22 &28.429&$<$80.0& 0.01 &22.699&....&14,19,4,1,4,4\\
NGC7582     & 0.005 & 5.99 & 7.48 & 52.47& 83.27& 148.0 &  ...    &  ...   & 41.63 &29.317&155.0& 0.26 & 23.95& 521&3,3,1,3,1,11,4,4\\
NGC7590     & 0.005 & 6.83 & 0.89 & 7.69 & 20.79&   7.8 &    1.5  & 5.6    & 40.02 &28.645& 12.0&0.006&$<$21.0&....&7,14,19,4,1,11,4\\
NGC7672     & 0.013 & 6.82 &$<$0.15&0.46&$<$2.46& ...   &   ...   &   ...  & ...   &28.382&286.0& ...  & 25.00&....&3,3,5,4,7\\
NGC7679     & 0.017 & 8.56 & 1.12 & 7.40 & 10.71&  ...  & ...     &    ... & 41.78 &29.708$^a$& 45.8 & 0.37 & 20.34&$<$200&7,14,7,1,1,7,12\\
UGC6100     & 0.03  & 8.26 & 0.202& 0.574& 1.50 &  ...  &  ...    &  ...   & 42.30 &29.265&$<$114.0& 1.22 & ...  &....&3,1,3,5,4\\
\hline

\end{tabular}

\end{small}

\end{table*}

\begin{table*}
\caption{\small The HBLR Sy2 sample}
\begin{small}
\setlength{\tabcolsep}{1.8pt}
\renewcommand{\arraystretch}{0.6}
\begin{tabular}{lccccccccccccccl} \hline \hline
Name & $z$ & $M_{\rm BH}$ & $f_{25}$ & $f_{60}$ & $f_{100}$ & \Neii   & \Nev & \oiv &$L_{\rm [O~ III]}$ & $ L_{\rm 1.49 GHz}$ &$F_{\rm HX}$ & $\dot{M}$ &$ N_{\rm H}$& EW & Reference \\
      (1) & (2) &(3) & (4) & (5) & (6)&(7)& (8) & (9)& (10)& (11)&(12) &(13)& (14) &(15)&(16)\\
\hline

Circinus      & 0.001 & ...  & 68.44 & 248.7 &315.85&   900.0& 239.0  & ...  & 40.92 & ...  & 100.0& 0.05 & 24.633&2250&1,1,4,11,4,4\\
ESO273-IG04   & 0.039 & ...  & 1.72  & 4.76  & 4.92 &   ...  & ...    &  ... & 42.48 & ...  & ...  & 1.85 & ...   &....&1,4\\
F00317-2142   & 0.027 & 8.08 & 0.56  & 3.85  & 8.42 &  ...   &  ...   & ...  & 41.13 &30.005$^a$& ...  & 0.08 & 20.28 &$<$900&7,1,7,1,7,12\\
F00521-7054   & 0.069 & ...  & 0.80  & 1.02 &$<$1.44& 5.8    & 5.78   &  8.63& 42.62 & ...  &$<$318.0& 2.55 & ...   &....&1,19,4,4\\
F01475-0740   & 0.018 & 7.55 & 0.84  & 1.10  & 1.05 &  16.0  &  ...   &  7.0 & 41.76 &30.278& 7.50 & 0.35 & 21.59 & 130&9,2,17,4,5,4,4,4\\
F02581-1136   & 0.03  & ...  & 0.46  & 0.54  & 0.85 &  ...   &  ...   &  ... & 41.16 &29.175& ...  & 0.09 & ...   &....&2,4,5\\
F04385-0828   & 0.015 & 8.77 & 1.70  & 2.91  & 3.55 & 24.0   &   ...  & 12.0 & 40.64 &28.882& 24.0 & 0.027& ...   &....&9,2,17,4,5,4\\
F05189-2524   & 0.043 & 7.86 & 3.41  & 13.27 & 11.90&  21.12 & 17.53  & 23.71& 42.74 &29.999& 43.0 & 3.37 & 22.756&  30&7,2,18,4,5,11,4,4\\
F11057-1131   & 0.055 & ...  & 0.32  & 0.77  & 0.79 &  ...   &  ...   & ...  & 42.45 &29.592& 3.90 & 1.73 &$>$24.00&900 &1,4,5,4,4,4\\
F15480-0344   & 0.03  & 8.22 & 0.72  & 1.09  & 4.05 &  7.0   &  10.0  & 34.0 & 43.02 &29.863& 3.70 & 6.41 &$>$24.20&$<$2400&9,2,17,4,5,4,4,4\\
F17345+1124   & 0.162 & ...  & 0.20  & 0.48  & 3.31 &  ...   &   ...  &  ... & 42.956&31.782& ...  & 5.53 & ...   &....&1,5,5\\
F18325-5926   & 0.02  & ...  & 1.39  & 3.23  & 3.91 & ...    &  ...   &  ... & 42.19 & ...  & 84.0 & 0.95 & 22.31 &242 &1,4,11,4,4\\
F20050-1117   & 0.031 & 7.11 & 0.17  & 1.11  & 2.00 &  ...   &  ...   &  ... & 41.47 & ...  & ...  & 0.18 &$<$21.60&272&7,1,7,7,12\\
F20460+1925   & 0.181 & ...  & 0.53  & 0.88  & 1.45 &  ...   &  ...   &  ... & 43.02 &31.064& 15.0 & 6.41 & 22.398& 260&1,4,5,11,4,4\\
F22017+0319   & 0.061 & ...  & 0.59  & 1.31  & 1.65 &   5.95 &   8.33 & 29.04& 42.58 &30.060& 36.0 & 2.33 & 22.69 & 380&2,19,4,5,4,4,4\\
F23060+0505   & 0.173 & ...  & 0.43  & 1.15  & 0.83 &  ...   &  ...   & ...  & 43.916&30.584& 15.0 & 50.48& 22.924& 170&1,5,5,11,5,11\\
IC 1631       & 0.031 & 7.78 & 0.13  & 1.05  & 2.43 &  ...   &  ...   & ...  & 41.98 &29.591$^a$&100.0& 0.58 &$<$21.5&$<$70&7,1,7,1,11,7,11\\
IC 3639       & 0.011 & 6.83 & 2.54  & 8.90  & 13.79&  ...   &  ...   & ...  & 41.89 &29.265& 0.80 & 0.48 &$>$24.2&1500&9,14,4,5,4,4,4\\
IC 5063       & 0.011 & 7.20 & 3.95  & 5.79  & 3.66 &  21.0  &  ...   &  ... & 41.56 & ...  &120.0 & 0.22 & 23.342&  80&7,2,1,4,11,4,2\\
MCG-3-34-64   & 0.017 & 7.69 & 2.88  & 6.22  & 6.37 &  ...   &  ...   &  ... & 42.39 &30.158& 40.0 & 1.50 & 23.61 & 200&9,2,4,5,4,4,2\\
MCG-3-58-7    & 0.032 & ...  & 0.98  & 2.60  & 3.62 &    5.0 &  ...   & 12.0 & 41.93 &29.374& ...  & 0.52 & ...   &....&2,17,4,5\\
MCG-5-23-16   & 0.008 & 6.29 & ...   & ...   & ...  &   ...  &   ...  &  ... & 41.81 & ...  & 730.0& 0.40& 22.25  &35.2&7,4,11,4,4\\
Mrk 3         & 0.014 & 8.50 & 2.90  & 3.77  & 3.36 &   86.0 & 109.0  & 210.0& 43.27 &30.600& 65.0 &11.40& 24.134 & 610&7,1,17,4,5,11,4,4\\
Mrk 78        & 0.037 & 7.99 & 0.56  & 1.11  & 1.13 &   ...  &  ...   &  ... & 41.98 & ...  & ...  & 0.58 & ...   &....&7,1,7\\
Mrk 348       & 0.015 & 7.18 & 1.02  & 1.43  & 1.43 &   13.0 &  ...   & 24.0 & 41.96 &30.132& 127.0& 0.56 & 23.204& 230&7,2,17,4,5,11,4,2\\
Mrk 463E      & 0.051 & 7.88 & 1.49  & 2.21  & 1.87 &   ...  & 18.3   & ...  & 42.89 &31.272&4.0   & 4.75 & 23.20 & 340&9,2,1,4,5,11,2,4\\
Mrk 477       & 0.038 & 7.20 & 0.51  & 1.31  & 1.85 &   ...  &   ...  &  ... & 43.62 &30.230& 12.0 & 25.53&$>$24.0& 490&9,10,4,5,11,4,4\\
Mrk 1210      & 0.014 & ...  & 2.08  & 1.89  & 1.30 &  ...   &  ...   & ...  & 42.37 &29.571& 13.0 & 1.44 & 23.263& 108&1,4,5,11,4,4\\
NGC 424       & 0.012 & 7.78 & 1.76  & 2.00  & 1.74 &   8.7  &  16.1  & 25.8 & 41.56 &28.752& 16.0 & 0.22 & 24.00 & 790&7,2,19,4,5,4,2,4\\
NGC 513       & 0.02  & 6.29 & 0.48  & 0.41  & 1.32 &  12.76 &   1.91 &  6.54& 41.14 &29.621& ...  & 0.08 & ...   &....&7,2,19,4,5\\
NGC 591       & 0.015 & 6.84 & 0.448 & 1.99  & 3.48 &  ...   &  ...   &  ... & 41.97 &29.187& 2.0  & 0.57 &$>$24.2&2200&7,1,4,5,4,4,4\\
NGC 788       & 0.014 & 7.51 & 0.51  & 0.51  & 0.59 &  ...   &  ...   & ...  & 40.79 & ...  & 46.20& 0.038& 23.324&....&6,1,4,4,4\\
NGC 1068      & 0.004 & 7.64 & 92.7  & 198.0 & 259.8&  520.0  & 1110.0 &2200.0& 42.33 &30.130& 35.0& 1.31 & 25.00 &1210&9,2,1,4,5,11,2,2\\
NGC 2110      & 0.008 & 7.96 & 0.84  & 4.13  & 5.68 &  ...   &   ...  &  ... & 41.35&29.770$^a$&260.0& 0.14  &22.17 & 124&7,1,7,1,11,7,11\\
NGC 2273      & 0.006 & 7.30 & 1.36  & 6.41  & 9.55 &   ...  &  16.08 &  ... & 41.13 &28.803& 9.0  & 0.08 & 24.13 &2200&6,14,1,4,5,11,4,4\\
NGC 2992      & 0.008 & 7.72 & 1.57  & 7.34  & 11.6 &  ...   &  ...   &  ... & 41.30 & ...  & 45.0 & 0.12 & 21.84 & 514&9,13,4,11,4,4\\
NGC 3081      & 0.008 & 6.29 & ...   & ...   & ...  &  ...   &  ...   &  ... & 41.43 &27.731& 13.0 & 0.16 & 23.819& 610&7,4,5,11,4,4\\
NGC 3185      & 0.004 & 6.06 & 0.14  & 1.43  & 3.67 &   ...  & ...    &  ... & 39.85 &27.416$^a$&...&0.004&$\le$21.30&....&7,8,7,1,7\\
NGC 4388      & 0.008 & 7.22 & 3.72  & 10.46 & 18.1 &  54.0  &  56.0  &  ... & 41.85 &29.188& 120.0& 0.43 & 23.43 & 440&9,2,1,4,5,11,4,4\\
NGC 4507      & 0.012 & 6.42 & 1.39  & 4.31  & 5.40 &   ...  &  18.4 &  ... & 42.19 &29.190& 210.0 & 0.95 & 23.643& 117&7,1,1,4,5,11,4,4\\
NGC 5252      & 0.023 & 8.04 & ...   & ...   & ...  &  ...   &  ...   &  ... & 42.05 &29.212& 89.0 & 0.69 & 22.461&  44&9,4,5,11,4,4\\
NGC 5506      & 0.006 & 7.46 & 4.24  & 8.44  & 9.24 &  59.0  &  82.0  & ...  & 41.45 &29.360& 838.0& 0.17 & 22.46 & 150&7,2,1,4,5,11,4,2\\
NGC 5995      & 0.024 & 7.11 & 1.45  & 4.09  & 7.06 &  ...   &  ...   &  ... & 42.98 &29.561& 220.0& 5.85 & 21.934& 240&7,2,4,5,4,4,2\\
NGC 6552      & 0.027 & ...  & 1.17  & 2.57  & 2.79 &  ...   &  ...   & ...  & 42.41 &29.640& 6.00 & 1.57 & 23.80 & 900&2,4,5,11,2,2\\
NGC 7212      & 0.027 & 7.48 & 0.77  & 2.89  &4.90  &  ...   &  ...   & ...  & 42.73 &30.188& 6.90 & 3.29 &$>$24.2& 900&7,1,4,5,4,4,4\\
NGC 7314      & 0.005 & ...  & 0.58  & 3.74  & 1.42 &  ...   &  23.0  & 53.0 & 42.41 & ...  &356.0 & 1.57 & 22.02 & 147&1,19,4,11,4,4\\
NGC 7674      & 0.029 & 7.56 & 1.79  & 5.64  & 8.46 &   18.0 &  31.0  & 46.0 & 42.57 &30.572& 5.00 & 2.28 &$>$24.00 & 370&9,2,17,4,5,11,4,4\\
NGC 7682      & 0.017 & 7.28 & 0.22  & 0.47  & 0.41 &   ...  &  ...   &  ... & 41.76 &29.545&$<$130.0& 0.35 & ...   &....&9,2,4,5,4\\
Was 49b       & 0.063 & ...  & ...   & ...   & ...  &  ...   &  ...   &  ... & 42.51 &30.755& 6.30 & 1.98 & 22.799& 620&4,5,4,4,4\\
\hline

\end{tabular}
{ \noindent \vglue 0.5cm {\sc Notes}: Col. (1): Source name; Col.
(2): Redshift; Col. (3): Log of SMBHs masses in units of $M_{\sun}$;
Col. (4), (5), and (6): Infrared flux (in Janskys) for 25, 60, and
100 $\mu m$; Col. (7), (8), and (9): Flux ($10^{-21}$W $\rm
cm^{-2}$) for \Neii $\lambda 12.81$,\Nev $\lambda 14.32$, and \oiv
$\lambda 25.89$; Col. (10): Log of extinction-corrected
\oiii$\lambda$5007 luminosity in units of ergs $\rm s^{-1}$; Col.
(11): Log of luminosity of radio for 1.49 GHz in ergs $\rm s^{-1}$
$\rm {Hz}^{-1}$; Col. (12): Fluxes of observed X-ray (2-10 keV) in
units of $10^{-13} \rm ergs ~s^{-1} cm^{-2}$ for Sy2s; Col. (13):
Accretion rates ($M_{\sun}$ $\rm yr^{-1}$);
 Col. (14): Log of gaseous absorbing column
density ($N_{\rm H}$) in units of $\rm cm^{-2}$;  Col. (15): EW is
the Fe k$\alpha$ equivalent width in eV. Col. (16): References (for
cols. [3], [4]-[6], [7]-[9], [10], [11], [12], [14], and [15],
respectively).

$^a$$L_{\rm 1.49GHz}$ are the luminosities at 1.4 GHz.

$^b$19 are references for 19, 1, and 19, respectively.

$^c$19 is reference for 19 and 1, respectively.

$^d$EW are the equivalent width of Fe K$\alpha$ line at 6.7 keV.

{\sc References}: (1) NED; (2) Tran 2003; (3) Zhang \& Wang 2006;
(4) Shu et al. 2007; (5) Gu \& Huang 2002; (6) Bian \& Gu 2007; (7)
Wang \& Zhang 2007; (8) Ho et al. 1997; (9) Wang et al. 2007; (10)
Imanishi, M 2002; (11) Bassani et al. 1999; (12) Panessa \& Bassani
2002; (13) Surace et al. 2004; (14) Sanders et al. 2003; (15) Teng
et al. 2005; (16) Akylas \& Georgantopolos 2009; (17) Deo et al.
2007; (18) Farrah et al. 2007; (19) Tommasin et al. 2008; (20)
Goulding \& Alexander 2009; (21) Marconi et al. 2001.}
\end{small}
\end{table*}

To present the properties and the dominant mechanisms between the
two groups, we calculate some parameters, for example, far-infrared,
infrared, 1.49 GHz, \oiii luminosities, and high excitation lines
ratios (\Nev/\Neii and \oiv/\Neii) and introduce some of them as
follows.

In order to get more luminosities of different bands, we need to
calculate the luminosity distances of some objects. The luminosity
distance can be shown as

$$D_{\rm L}=(1+z)\frac{c}{H_0}\int^z_0
[(1+x)^3\Omega_{\rm M}+\Omega_{\Lambda}]^{-0.5}dx$$ (Darling \&
Giovanelli 2002; Ballantyne et al. 2006).

We calculate the luminosities of far-infrared and infrared bands of
most of sources in our sample by the fluxes from either the
published papers or the NASA/IPAC Extragalactic Database (NED). The
fluxes and luminosities of far-infrared and infrared can be shown to
be: $ F_{\rm fir}=1.26\times10^{-14}(2.58 f_{\rm 60}+f_{\rm
100})[\rm Wm^{-2}]$, $L_{\rm fir}=4\pi{\rm C}{D_{\rm L}}^2 F_{\rm
fir}[L_{\sun}]$, $ F_{\rm ir}=1.8\times10^{-14}(13.48 f_{12}+5.16
f_{25}+2.58f_{60}+f_{100})[\rm Wm^{-2}]$, \textbf{and} $ L_{\rm
ir}=4\pi {D_{\rm L}}^2  F_{\rm ir}[L_{\sun}]$, where the constant C
is the correction factor required to account principally for the
extrapolated flux longer than the Infrared Astronomical Satellite
($IRAS$) 100 $\mu m$ filter (Sanders \& Mirable 1996), and here
C=1.6; the fluxes for 25, 60, and 100 $\mu$m are listed in Tables 1
and 2, while the 12 $\mu$m fluxes (not appearing in Tables 1 and 2
for simplicity) are also selected from the same literatures or NED
as the 25, 60, 100 $\mu$m fluxes.

Besides the luminosities of radio band (1.49 GHz) of some sources
from literatures, we also obtain the luminosities of other sources
using $ L=4\pi D_{\rm L}^2 F$, where $L$ and $F$ are the luminosity
and flux of the radio band (1.49 GHz) from NED. The 2-10 keV fluxes
come directly from some published literatures.

The \oiii luminosity could be taken as an indicator of the nuclear
activity only after correction for extinction (Maiolino et al. 1998;
Bassani et al. 1999; Gu \& Huang 2002). The luminosity of the
extinction-corrected \oiii $\lambda 5007$ emission is given as $
L_{\rm [O~III]} = 4 \pi D_{\rm L}^{2} F_{\rm [O~III]}^{\rm cor}$,
where $ F_{\rm [O~III]}^{\rm cor}$ is the extinction-corrected flux
of \oiii$\lambda 5007$ emission derived from the relation (Bassani
et al. 1999)
$$ F_{\rm [O~III]}^{\rm cor} = F_{\rm [O~III]}^{\rm obs} \rm
[\frac{(H\alpha/H\beta)_{obs}}{(H\alpha/H\beta)_{0}}]^{2.94}
$$

\noindent where an intrinsic Balmer decrement $\rm
(H\alpha/H\beta)_{0} = 3.0$ is adopted.

\begin{figure*}
\begin{center}
\includegraphics[width=8cm,height=10cm]{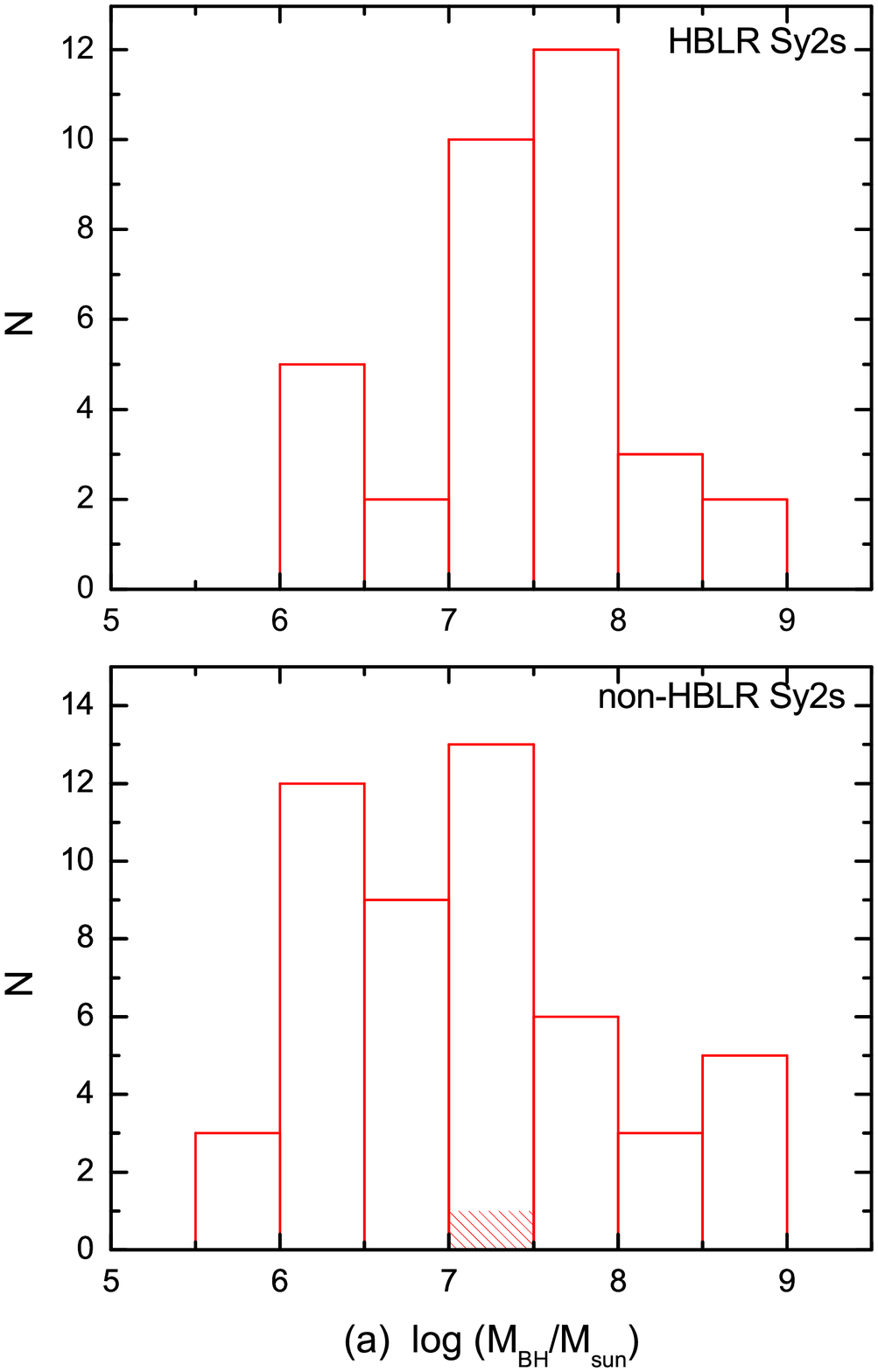}
\includegraphics[width=8cm,height=10cm]{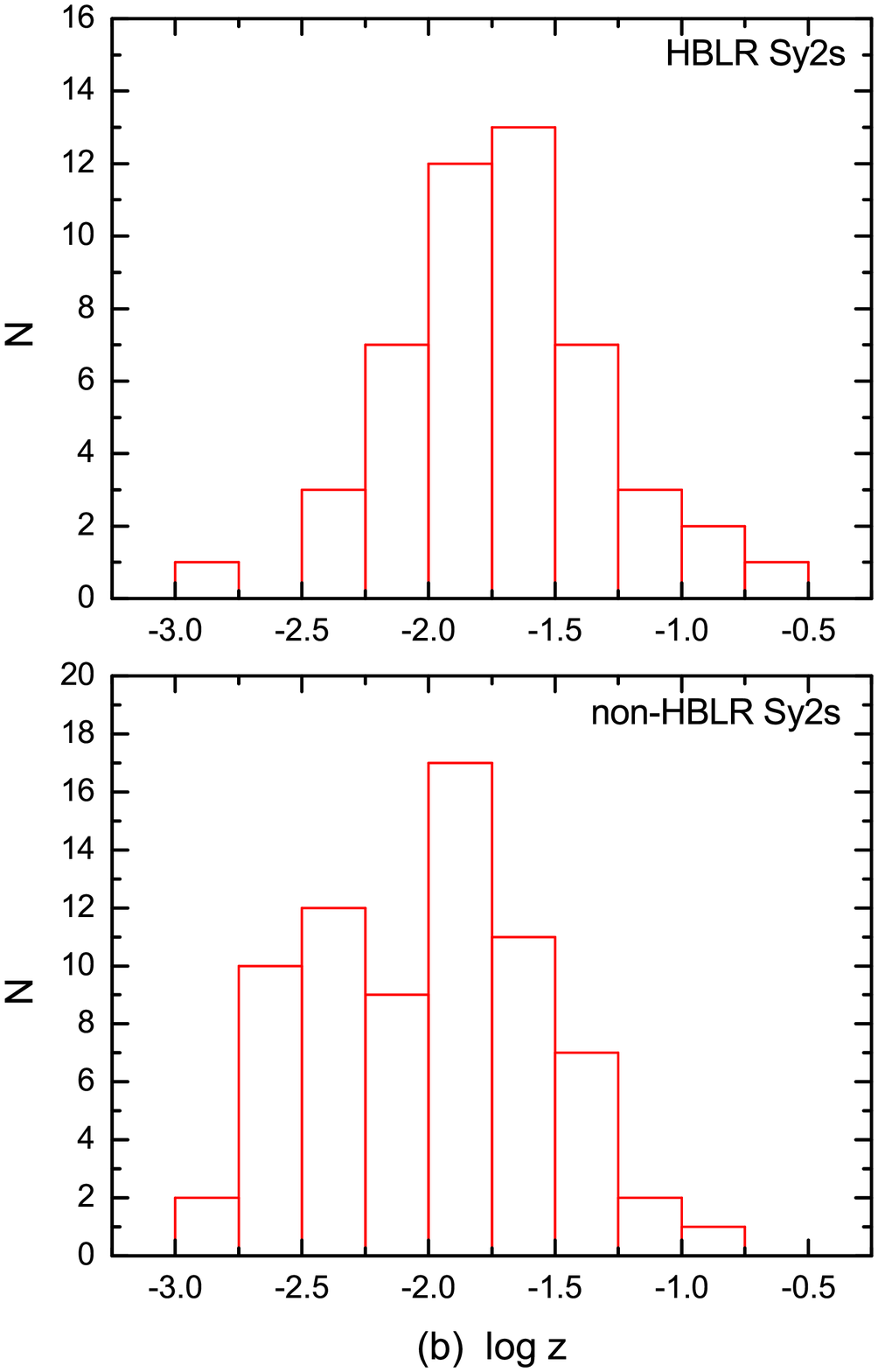}
\includegraphics[width=8cm,height=10cm]{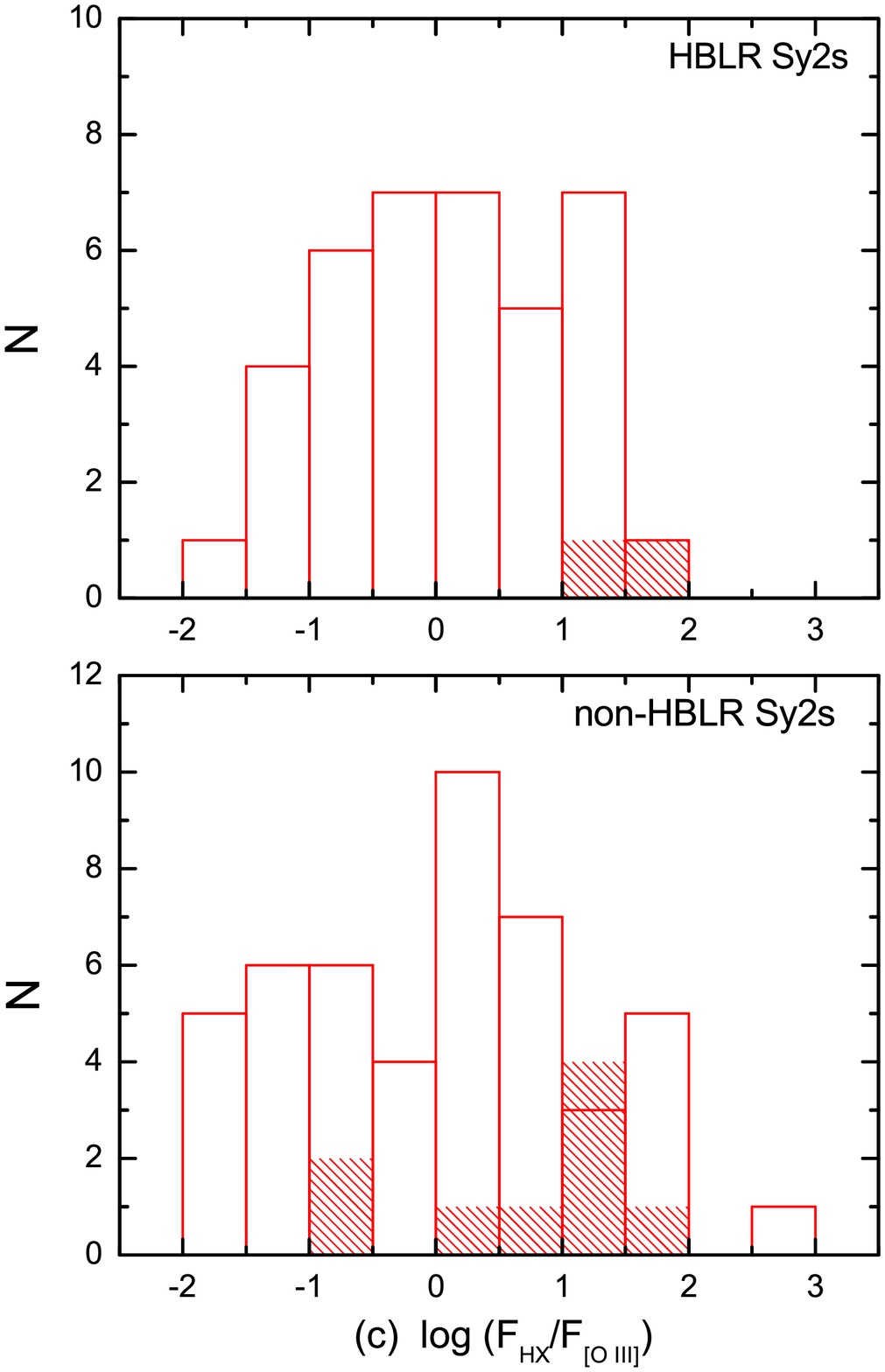}
\includegraphics[width=8cm,height=10cm]{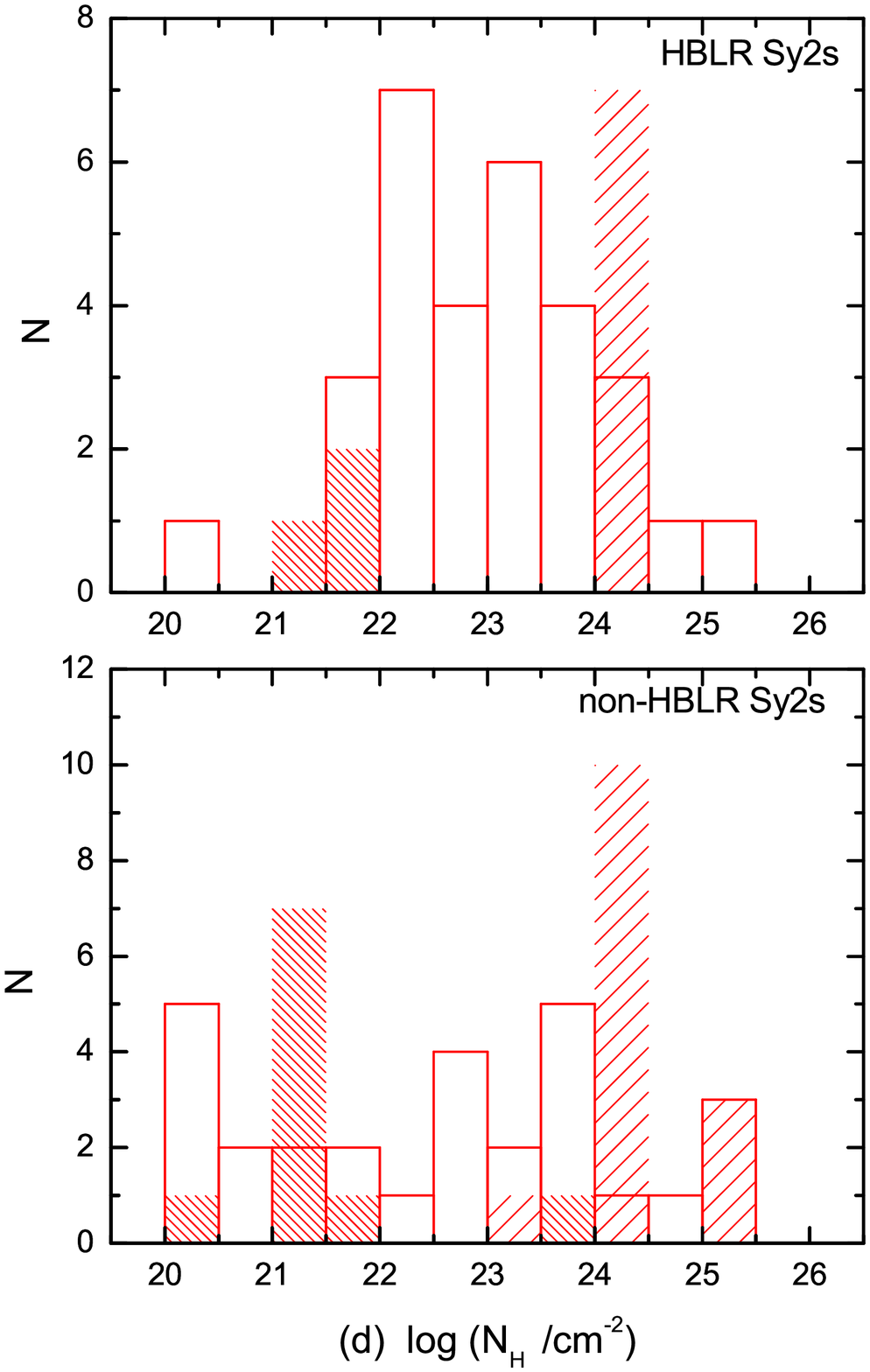}

\caption{Distributions of the mass of the black hole, redshift,
 $F_{2-10 \rm keV}/F_{\rm [O III]}$ ratio, and column density of neutral hydrogen.
Densely and sparsely shaded areas denote the upper and lower limits,
respectively.}
\end{center}
\end{figure*}

In addition to above mentioned quantities, we need a good diagnostic
that is sensible for the observational differences between the two
types of objects. We have chosen the lines of \Nev and \oiv because
they are not affected by photoionization of stars and because they
are generally among the brightest highly-ionized lines (Sturm et al.
2002). For examining starburst and AGN activities, the \Nev $\lambda
14.32$ and \oiv $\lambda 25.89$ lines are the most useful single
line diagnostics. Both lines are strong in spectra of AGNs (Farrah
et al. 2007), while they are weak in spectra of star-forming regions
(Lutz et al. 1998). We consider diagnostics based on the
fine-structure line ratios. In Tables 1 and 2, we list various
parameters both types of objects.

\begin{table*}
\caption{Summary of HBLR and non-HBLR Sy2s.}
\begin{small}
\begin{center}
\renewcommand{\arraystretch}{0.5}
\begin{tabular}{lccccl}
\hline \hline Parameters &\multicolumn{2}{c}{non-HBLR Sy2s}& $p_{\rm
null}$
& \multicolumn{2}{c}{HBLR Sy2s}\\
\cline{2-3}\cline{5-6} & Mean & N&(\%)
 & Mean &   N
 \\
(1)& (2) & (3) & (4) &(5)&(6)  \\
\hline
log($N_{\rm H})$                                     &22.96 $\pm$0.25      & 52&53.66& 23.18$\pm$0.19  & 40         \\
EW(Fe)$^{\rm a}$                                    &1325$\pm \alpha ^{\rm d}$ & 32 &1.83 & 544$\pm$99    & 38     \\
$F_{\rm HX}/ F_{[\rm O~ III]}$                      &24.34 $\pm$17.49      & 56  &28.15&6.02$\pm$1.55        & 40   \\
log$M_{\rm BH}$                                     & 7.10 $\pm$0.11      & 52 &3.59 & 7.41$\pm$0.11        & 34   \\
log z                                               &-1.98$\pm$0.05       & 71 &2.23 & -1.71$\pm$0.06       & 49   \\
log$L_{\rm 1.49GHz}$                                &29.01$\pm$0.13       & 56 &0.068&29.74$\pm$0.14        & 38   \\
$L_{\rm FIR}(10^{10}L_{\sun})$                      &6.21$\pm$1.60        & 64 &47.70& 9.22$\pm$3.05        & 44   \\
$L_{\rm IR}(10^{11}L_{\sun})^{\rm b}$               & 1.53$\pm$0.41       & 57 &17.19& 3.01$\pm$1.07        & 43   \\
${f_{60}/f_{25}}^{\rm c}$                           &5.81$\pm$0.45        &60  &$10^{\rm -4}$&2.99$\pm$0.30  & 45  \\
$\dot{M}$                                           &0.76$\pm$0.22        & 66 &0.01 & 3.11$\pm$1.14        & 49   \\
$(F_{\rm [Ne~V]}/F_{\rm [Ne~II]})$                  &0.40$\pm$0.12        & 21 &0.17 &1.21$\pm$0.17         & 12   \\
$(F_{\rm [O~ IV]}/F_{\rm [Ne~II]})$                 &1.03$\pm$0.27        &23  &0.72 &2.33$\pm$0.44         & 13   \\
log$L_{\rm [O~ III]}$                               &41.05$\pm$0.17       & 66 &0.01 &42.05$\pm$0.11        & 49   \\
log($L_{\rm bol}/L_{\rm Edd})^{\rm e}$              &-0.98$\pm$0.25       & 49 &5.75 &-0.12$\pm$0.15        & 34   \\

\hline \hline
\end{tabular}
\parbox{6.5in}
{\baselineskip 9pt \noindent \vglue 0.5cm {\sc Note}: Col.(1):
Parameters; Cols.(2)-(3) and (5)-(6): For each sample of non-HBLR
Sy2 and HBLR Sy2 galaxies, {``Mean''} is the mean value of the
various parameters and N is the number of data points. Col. (4): the
probability $p_{\rm null}$ (in percent) for the null hypothesis that
the two distributions are drawn at random from the same parent
population. When there are censored data, we use Gehan's generalized
Wilcoxon test (hypergeometric variance) in ASURV.

$^a$ EW(Fe) is the Fe K$\alpha$ equivalent width in eV.

$^b$ We have removed 3 sources (NGC 1241, NGC 3362, and NGC 7682)
because they have no detections in their 12 $\mu m$ band.

$^c$ Detections only.

$^d$ An ASURV test does not give the value of $\alpha $.

$^e$ Here the Eddington ratios, $L_{\rm bol}/L_{\rm Edd}$ are given
by $L_{bol}=3500L_{\rm [O~III]}$ and $L_{\rm
Edd}=1.4\times10^{38}(M_{\rm BH}/M_{\sun})\rm ergs~s^{-1}$,
respectively.}

\end{center}
\end{small}
\end{table*}

\begin{figure*}
\begin{center}
\includegraphics[width=12cm,height=8cm]{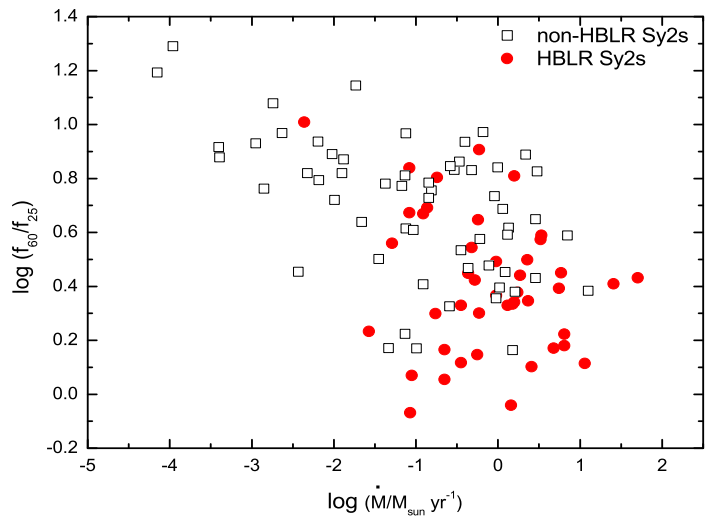}
\caption{IR color $f_{60}/f_{25}$ ratio vs. accretion rate (defined
as $\dot{M}=L_{\rm bol}/{\eta c^2}$) for HBLR and non-HBLR Sy2s,
where $\eta$=0.1 is the accretion efficiency (Wang et al. 2007); the
filled circles denote HBLR Sy2s and the open squares denote non-HBLR
Sy2s.}
\end{center}
\end{figure*}

\section{RESULTS AND DISCUSSION}
In sections 3.1 and 3.2, we will show the properties that differ
between the two groups and test if HBLR Sy2s are dominated by AGNs,
and if non-HBLR Sy2s are dominated by starbursts. In section 3.3, we
employ the separation of Shu et al. (2007) who noted that non-HBLR
Sy2s are divided into the luminous ($L_{\rm [O~III]}>10^{41}
\rm~ergs~s^{-1}$) and less luminous ($L_{\rm [O~III]}<10^{41}
\rm~ergs~s^{-1}$) classes. We will investigate their differences in
the obscuration between the two groups. We also discuss their
properties and compare their obscuration with that of HBLR Sy2s.

\begin{figure*}
\begin{center}

\includegraphics[width=8.5cm,height=6cm]{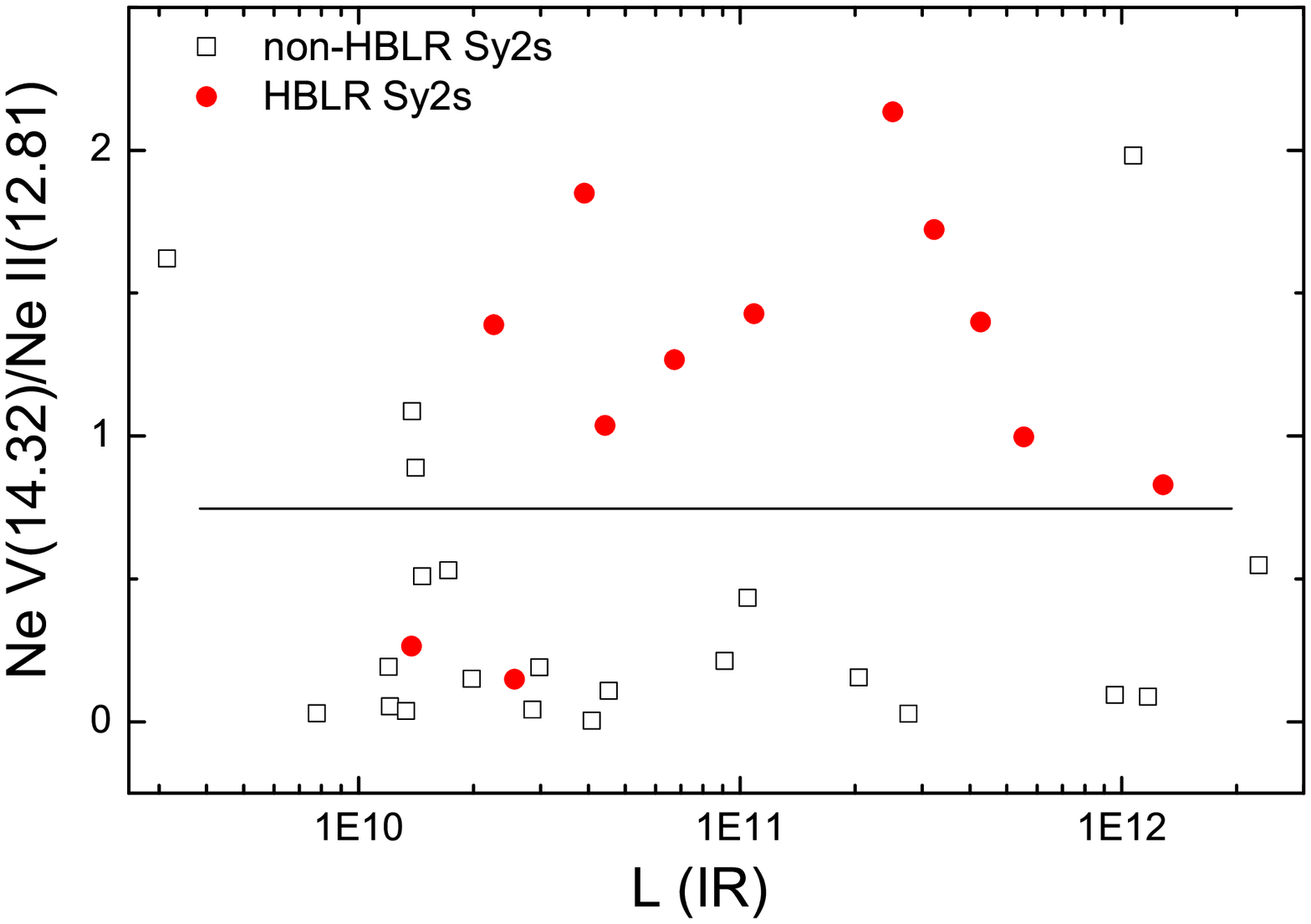}
\includegraphics[width=8.5cm,height=6cm]{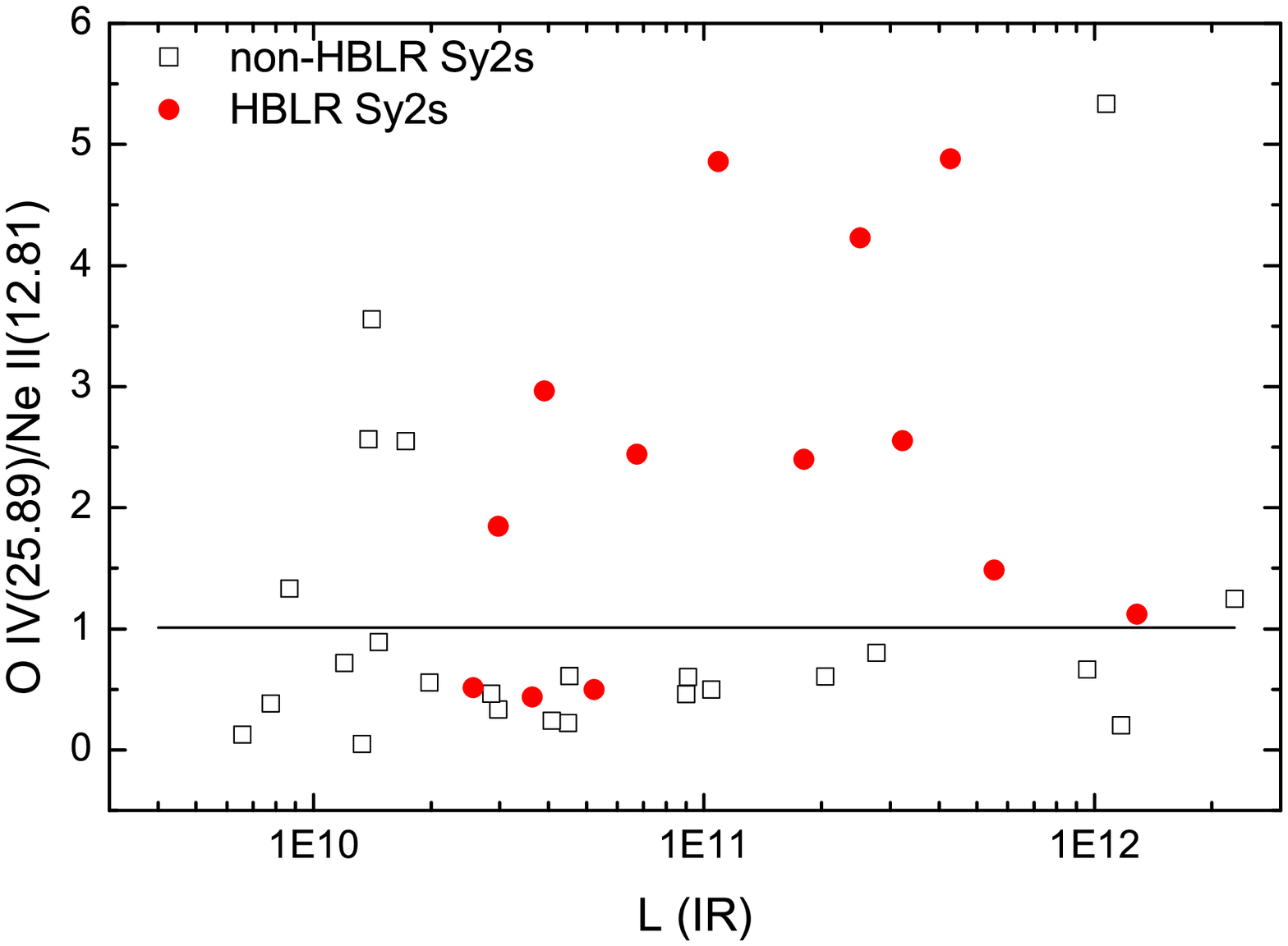}
\caption{IR luminosity L(IR) vs. \Nev $\lambda 14.32$/\Neii $\lambda
12.81$ ratio (left) and \oiv $\lambda 25.89$/\Neii $\lambda 12.81$
ratio (right). The solid line probably shows the starburst or AGN
dominated region. The filled circles denote HBLR Sy2s and the open
squares denote non-HBLR Sy2s.}
\end{center}
\end{figure*}

\subsection{Distributions of Main properties for Non-HBLR and HBLR Sy2s}

In this section, we report the distributions of several parameters
for HBLR and non-HBLR Sy2s, for example, the SMBH mass, redshift,
$N_{\rm H}$, $F_{\rm 2-10keV}/F_{[\rm O~III]}~(F_{\rm HX}/F_{[\rm
O~III]})$~ratio, K$\alpha$ iron-line equivalent width (EW), and
mid-infrared line ratios, both $F_{\rm [Ne~V]}/F_{\rm [Ne~II]}$ and
$F_{\rm [O~IV]}/F_{\rm [Ne~II]}$.

 In Figure 1.a, we show the distribution of
SMBH masses for HBLR and non-HBLR Sy2s. Since there are censored
data points (upper limits; densely shaded areas) among non-HBLR
Sy2s, we use the astronomical survival analysis package (ASURV;
Feigelson $\&$ Nelson 1985) for statistical analysis. The
distributions of the SMBH masses are different between HBLR and
non-HBLR Sy2s. For the whole sample, the mean SMBH mass of HBLR Sy2s
is larger than that of non-HBLR Sy2s by the amount of 0.31, with a
confidence level of $96.41\%$ (see Table 3).

Figure 1.b shows the distribution of redshifts for HBLR and non-HBLR
Sy2s. The distributions of redshifts are different between HBLR and
non-HBLR Sy2s. The mean value of log z of HBLR Sy2s is larger than
that of non-HBLR Sy2s by the amount of $0.27$. A Kolmogorov-Smirnov
(K-S) test shows that the probability for the two samples to be
extracted from the same parent population is $2.23\%$.

In Figure 1.c, $F_{\rm HX}/F_{[\rm O~III]}$, which is the ratio
{``$T$''}, is a good indicator of obscuration (Bassani et al. 1999;
Tran 2003; here, \oiii fluxes have been corrected for extinction,
and X-ray fluxes have not been corrected for absorption). Table 3
shows little difference in $F_{\rm HX}/F_{[\rm O~III]}$ between the
two groups. An ASURV test shows that the probability for the two
samples to be extracted from the same parent population is about
$28.15\%$. The mean value of $T$ is $24.34\pm17.49$ for non-HBLR
Sy2s and $6.02\pm 1.55$ for HBLR Sy2s.

In Figure 1.d, the $N_{\rm H}$ distributions between HBLR and
non-HBLR Sy2s are not significantly different (with a confidence
level of $53.66\%$; see Table 3; since an ASURV test could not deal
with a case that contained both upper and lower limits, we adopt the
$N_{\rm H}$ upper limits as the measured values). Our $N_{\rm H}$
distribution is consistent with the results of Gu \& Huang (2002),
Tran (2003) and Shu et al. (2007). This may be explained by the
following reason: since the mean value of $N_{\rm H}$ is
$10^{21.85\pm0.33}\rm cm^{-2}$ for the less luminous non-HBLR Sy2s
(see Table 4), they weaken greatly the difference in $N_{\rm H}$
between non-HBLR Sy2s and HBLR S2ys (Shu et al. 2007). In section
3.3, we will find that $N_{\rm H}$ has the significant differences
among the luminous, less luminous non-HBLR Sy2s, and HBLR Sy2s (see
Table 4 and Figure 4).

In Table 3, non-HBLR Sy2s are obviously larger in terms of the mean
value of EW(Fe) than HBLR Sy2s. An ASURV test shows that the
difference between the two samples is present (at a level of
$98.17\%$). Our result is not consistent with that of Tran (2003) or
Shu et al. (2007). This may be explained by the following reason:
for the small sample size (only 11 objects) of the less luminous
non-HBLR Sy2s with EW(Fe) measurements, adding them to the non-HBLR
Sy2 sample cannot weaken the difference ($99.95\%$) in EW(Fe) found
in the luminous Sy2 sample (see Table 4; Shu et al. 2007).

The two mid-infrared line ratios, $F_{\rm [Ne~V]}/F_{\rm [Ne~II]}$
and $F_{\rm [O~IV]}/F_{\rm [Ne~II]}$, can better distinguish HBLR
from non-HBLR Sy2s (see Table 3). Table 3 shows the significant
differences in the two ratios between the two groups. A K-S test
displays that the probabilities for the two sample to be extracted
from the same parent population are $0.17\%$ and $0.72\%$,
respectively.

\begin{figure*}
\begin{center}
\includegraphics[width=8cm,height=10cm]{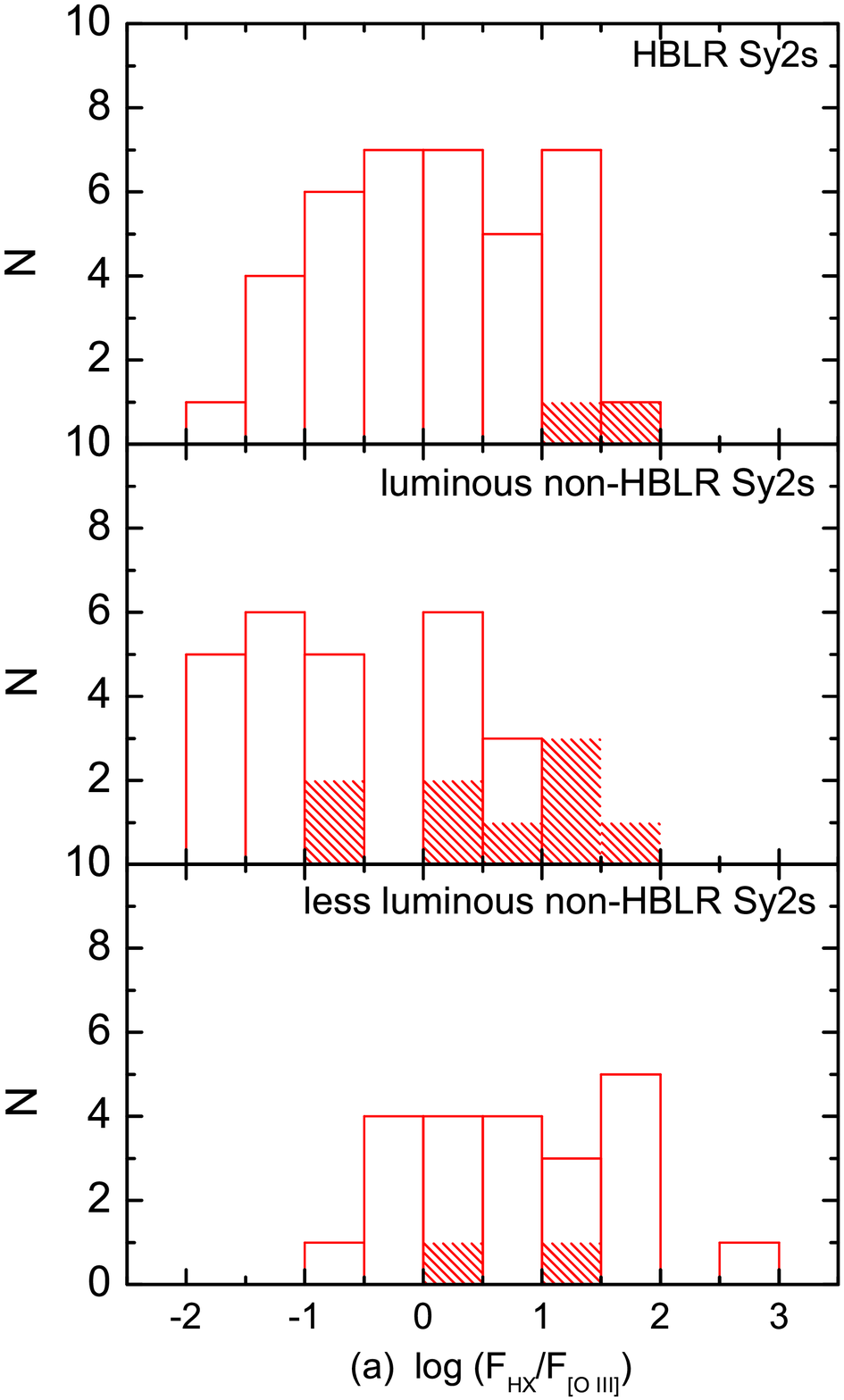}
\includegraphics[width=8cm,height=10cm]{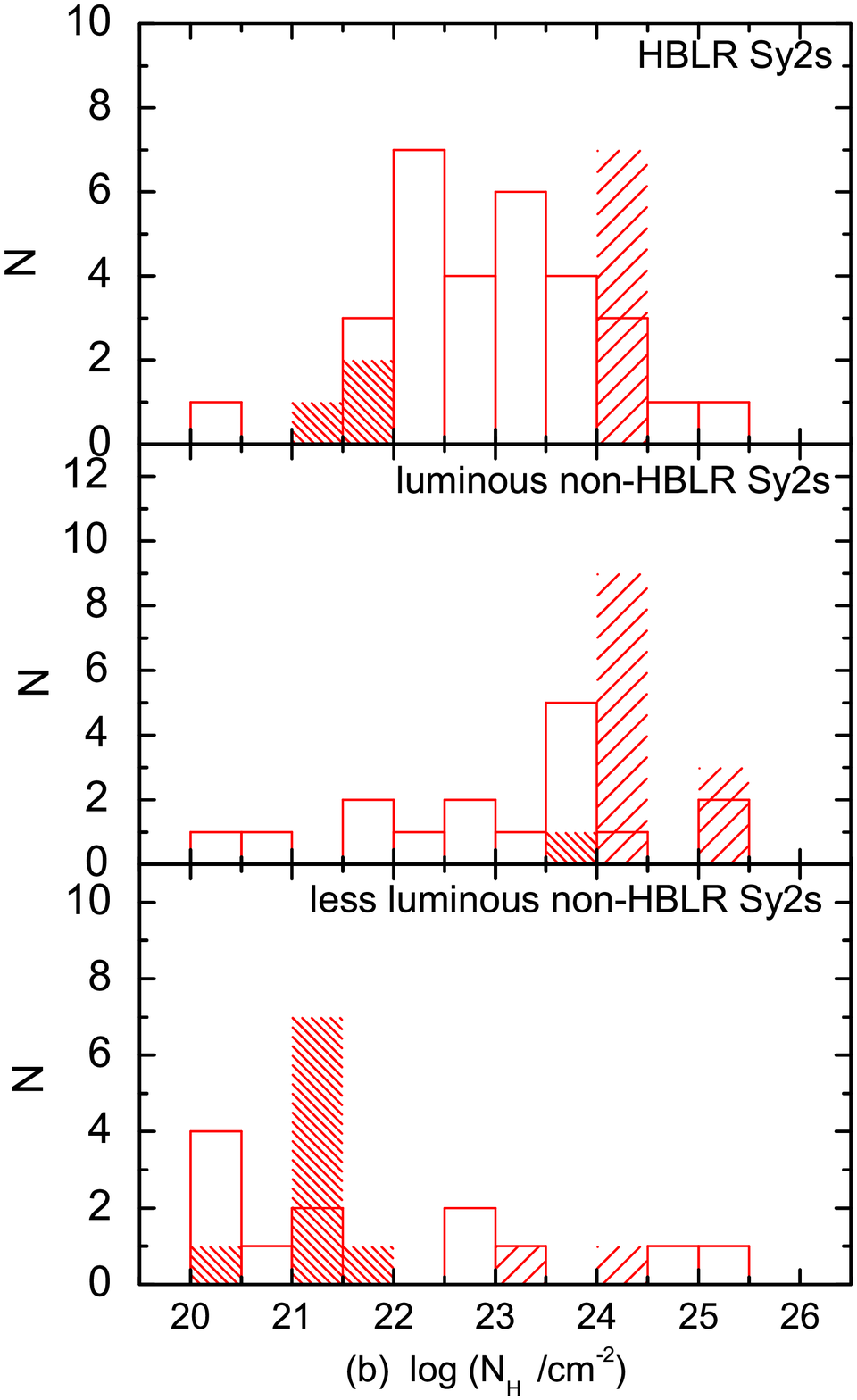}
\caption{Distributions of the $F_{2-10 \rm keV}/F_{\rm [O III]}$
ratio and column density of neutral hydrogen for the HBLR Sy2s,
luminous ($L_{\rm [O~III]}>10^{41} \rm~ergs~s^{-1}$) non-HBLR Sy2s,
and less luminous ($L_{\rm [O~III]}<10^{41} \rm~ergs~s^{-1}$)
non-HBLR Sy2s.
 Densely and sparsely shaded areas denote the upper and lower limits, respectively.}
\end{center}
\end{figure*}

In Table 3, we also provide other statistical results related to the
two types of sources. We find that most of the results of various
parameters show the significant differences between the two groups.
These indicate that HBLR and non-HBLR Sy2s are clearly different
subsamples.

\subsection{Starburst or AGN Domination in HBLR and Non-HBLR Sy2s}

In this section, we test if HBLR Sy2s are dominated by AGNs, and if
non-HBLR Sy2s are dominated by starbursts. Next we use two methods
to demonstrate different dominant mechanisms between non-HBLR and
HBLR Sy2s.

As mentioned in section 1, the AGN activity is the key to
understanding the differences between the two kinds of Sy2s. AGN
luminosity comes from the disk accretion onto central SMBHs. \oiii
$\lambda$5007 luminosity represents only an indirect (i.e.,
reprocessed) measurement of the nuclear activity, but
extinction-corrected $L_{[\rm O~ III]}$ is a good indicator of the
AGN activity for Type II AGNs (Kauffmann et al. 2003).

The $IRAS$ $f_{60}/f_{25}$ flux ratio is not a good indicator of the
inclination, but of the relative strength of the host galaxy and
nuclear emission (Alexander 2001; Shu et al. 2007). It has been
shown that HBLR Sy2s have smaller values of $f_{60}/f_{25}$ ratio,
compared to non-HBLR Sy2s (Heisler et al. 1997). In Table 3, the
mean value of the $f_{60}/f_{25}$ ratio is $5.81\pm 0.45$ for
non-HBLR Sy2s and $2.99\pm 0.30$ for HBLR Sy2s. The difference (at
the $99.999\%$ level) in color between them may be due to the
relative strength of the host galaxies and nuclear emissions
(Alexander 2001), and this is consistent with Baum et al. (2010) who
noted that the HBLR Sy2s have a higher ratio of AGN-to-starburst
contribution to the spectral energy distribution (SED) than non-HBLR
Sy2s, based on their distributions of several starburst and AGN
tracers. So, we can show that the $f_{60}/f_{25}$ ratio denotes the
relative strength of starburst and AGN emissions.

 In Figure 2, we
show the correlation of the $f_{60}/f_{25}$ ratio versus the
accretion rate (defined as $\dot M=L_{bol}/\eta c^2$), with
Pearson's correlation coefficient of $-0.54$ and a probability of
$<0.0001$ (hereafter, we exclude NGC 676, NGC 1143, NGC 1358, NGC
2685, NGC 4472, and NGC 4698, because they either do not be detected
or they are the upper limit of fluxes and below 1 Jy at $25 \mu m$
or $60 \mu m$.). These results show that they have a significant
anticorrelation. As the $f_{60}/f_{25}$ ratio drops, the $\dot M$
value increases. HBLR Sy2s show the smaller $f_{60}/f_{25}$ ratios
and larger $\dot M$ values, which may indicate a higher ratio of
AGN-to-starburst activity in the SED; non-HBLR Sy2s show the larger
$f_{60}/f_{25}$ ratios and smaller $\dot M$ values, which may
indicate a lower ratio of AGN-to-starburst activity in the SED.
Therefore, we suggest that the non-HBLR Sy2s are dominated by
starbursts, while the HBLR Sy2s are dominated by AGNs.

\begin{table*}
\caption{Summary of luminous, less luminous non-HBLR Sy2s and HBLR
Sy2s.}
\begin{small}
\begin{center}
\renewcommand{\arraystretch}{0.5}
\begin{tabular}{lccccccccl}
\hline \hline Parameters &\multicolumn{2}{c}{non-HBLR
Sy2sA(S1)$^{\rm a}$}
& \multicolumn{2}{c}{HBLR Sy2s(S2)}&\multicolumn{2}{c}{non-HBLR Sy2sB(S3)$^{\rm a}$}& \multicolumn{3}{c}{$p_{\rm null}(\%)$}\\
\cline{2-3} \cline{4-5} \cline{6-7}\cline{8-10} & Mean & N
 & Mean &   N& Mean &   N& S1-S2& S2-S3 & S1-S3
 \\
(1)& (2) & (3) & (4) &(5)&(6)&(7)&(8)& (9)& (10)   \\
\hline
log$N_{\rm H}$                                    &23.82 $\pm$0.26  &29 &23.18$\pm$0.19 & 40 &21.85 $\pm$0.33   & 23 & 3.02 &0.01 &0.01  \\
EW(Fe)$^{\rm b}$                         &1539$\pm \alpha^{\rm c}$  &21 &544$\pm$99    &38&955$\pm \alpha^{\rm c}$&11&0.05  &97.97&2.91 \\
$F_{\rm HX}/ F_{[\rm O~ III]}$                    &0.96 $\pm$0.31   &32 &6.02$\pm$1.55  & 40 &55.28$\pm$39.94   & 24 &0.04  &4.92 &0.01 \\
log$L_{\rm [O~ III]}$                             &41.90$\pm$0.09   &42 &42.05$\pm$0.11 & 49 &39.54$\pm$0.19    & 24 &20.55 &$10^{-11}$&$10^{-12}$\\
log($L_{\rm bol}/L_{\rm Edd})^{\rm d}$            &0.27$\pm$0.16    &28 &-0.12$\pm$0.15 & 34 &-2.57$\pm$0.27    & 21 &9.82  &0.01 &0.01 \\

\hline \hline
\end{tabular}
\parbox{6.5in}
{\baselineskip 9pt \noindent \vglue 0.5cm {\sc Note}: Col.(1):
Parameters. Cols.(2)-(3), (4)-(5), and (6)-(7): For each sample of
the non-HBLR Sy2sA, HBLR Sy2s, and non-HBLR Sy2sB, {``Mean''} is the
mean value of various parameters and {``N''} is the number of data
points. Col.(8): From the K-S or ASURV test of luminous non-HBLR
Sy2s (S1) vs HBLR Sy2s (S2), the probability $p_{\rm null}$ for the
null hypothesis that the two distributions are drawn at random from
the same parent population. Col.(9): As in col (8), but for HBLR
Sy2s (S2) vs less luminous non-HBLR Sy2s (S3). Col.(10): As in
col.(8), but for luminous non-HBLR Sy2s (S1) vs less luminous
non-HBLR Sy2s (S3). When there are censored data, we use Gehan's
generalized Wilcoxon test (hypergeometric variance) in ASURV.

$^a$ Non-HBLR Sy2sA and Sy2sB indicate the luminous ($L_{\rm
[O~III]}>10^{41} \rm~ergs~s^{-1}$) and less luminous ($L_{\rm
[O~III]}<10^{41} \rm~ergs~s^{-1}$) non-HBLR Sy2s, respectively.

$^b$ EW is the Fe K$\alpha$ equivalent width in eV.

$^c$ An ASURV test does not give the value of $\alpha $.

$^d$ Here the Eddington ratio is the same as Table 3.}

\end{center}
\end{small}
\end{table*}

We also use another diagnostic for examining starburst and AGN
activities. Due to the intense star formation in the nuclear region
of many active galaxies, some fraction of the measured fluxes of low
lying fine structure lines (excitation potential $\le50$ eV) will be
produced by photoionization from stars rather than AGNs, while the
high excitation lines (\oiv, \Nev) show little or no contamination
from possible starburst components (Sturm et al. 2002). Genzel et
al. (1998) found that \oiv/\Neii and \Nev/\Neii are much higher in
AGNs than in starbursts, which can now be confirmed on a broader
statistical basis. \oiv originates purely from the narrow line
region (NLR) in AGNs.  In a unified scheme, the NLR line luminosity
should be independently orientated and be a good tracer of AGN
power, in particular when using an extinction insensitive and modest
excitation line like \oiv (Sturm et al. 2002). Since the ionization
potential of \Nev $\lambda 14.32$ is $E_{ion}=97.1$ eV, \Nev is
unlikely to be strong in galaxies without an AGN (Voit 1992; Farrah
et al. 2007). While \Neii is a fairly good tracer of hot star
emission in starburst activity. In AGNs, the \Neii from the NLR is
more easily contaminated by starburst emission than the higher
excitation \oiv line (Sturm et al. 2002).

In Figure 3, the relations of infrared luminosity versus flux ratios
of \Nev $\lambda 14.32$/\Neii $\lambda 12.81$ and \oiv$\lambda
25.89$/\Neii$\lambda 12.81$ are shown. We can see in the two regions
of each plot that the upper region is primarily HBLR Sy2s and the
lower one is primarily non-HBLR Sy2s. The \Nev $\lambda 14.32$ and
\oiv $\lambda 25.89$ lines are the most useful single line
diagnostics (Farrah et al. 2007). As a result, we suggest that the
non-HBLR Sy2s are starburst-dominated, while the HBLR Sy2s are
AGN-dominated.

In Figures 2 and 3, we find that HBLR and non-HBLR Sy2s clearly show
the differences in the power sources. In Table 3, the differences in
the accretion rate ($\dot{M}=L_{\rm bol}/\eta c^2$), $f_{60}/f_{25}$
ratio, and two mid-infrared line ratio, $F_{\rm [Ne~V]}/F_{\rm
[Ne~II]}$ and $F_{\rm [O~IV]}/F_{\rm [Ne~II]}$, are significant. So
we hold that non-HBLR Sy2s are starburst-dominated, while HBLR Sy2s
are AGN-dominated.

\subsection {Physical Nature of the Various Obscuration}

In this section, we will investigate differences in the obscuration
and reasons of the absence of PBLs in luminous and less luminous
non-HBLR Sy2s, and compare them with those of HBLR Sy2s.

With regard to the obscuration between HBLR and non-HBLR Sy2s, our
result (see Table 3) and previous results (Tran 2003; Gu $\&$ Huang
2002; Shu et al. 2007) all show little difference. This reason is
that adding the less luminous Sy2s to the Sy2 sample weakens the
difference in obscuration found in the luminous Sy2 sample (Shu et
al. 2007). Next we discuss the possible physical explanations of the
absence of PBLs in the luminous and less luminous non-HBLR Sy2s,
respectively.

Table 4 shows that all differences in $N_{\rm H}$, EW(Fe), and
$F_{\rm HX}/F_{\rm [O~III]}$ between the luminous non-HBLR Sy2s and
HBLR Sy2s are significant. An ASURV test shows that the
probabilities for the two samples to be extracted from the same
parent population are $3.02\%$, $0.05\%$, and $0.04\%$,
respectively. These results suggest that luminous non-HBLR Sy2s show
larger obscuration than HBLR Sy2s. In addition, we do not find a
significant difference (with a confidence of level of $20.55\%$) in
$L_{[\rm O~III]}$ between luminous non-HBLR Sy2s and HBLR Sy2s, and
the mean values of log($L_{[\rm O~III]}/\rm ergs~s^{-1})$ are
$41.90\pm0.09$ and $42.05\pm0.11$, respectively. So we suggest that
the obscuration is the key cause that makes PBLs weaker or
nondetectable for the luminous non-HBLR Sy2s. Our explanation
supports Shu et al. (2007)'s suggestion that the absence of PBLs in
the luminous Sy2s arises from the obscuration. However, PBLs are not
detected in most (24/28) of the less luminous Sy2s in our sample.
The reason is still unclear.

To explore the natural reason of the absence of PBLs for less
luminous non-HBLR Sy2s, we analyse their obscuration: $N_{\rm
H}=10^{21.85\pm0.33}\rm~cm^{-2}$ and $F_{\rm HX}/F_{[\rm
O~III]}=55.28\pm39.94$ (see Table 4), suggesting that the less
luminous non-HBLR Sy2s have the smaller obscuration than the
luminous non-HBLR Sy2s or HBLR Sy2s\footnote{Because the sample size
of less luminous non-HBLR Sy2s with EW(Fe) measurements is only 11
and NGC 3982 has an EW(Fe) of 6310~eV, Table 4 shows almost no
difference in EW(Fe) between less luminous non-HBLR and HBLR Sy2s.
However, the differences in $N_{\rm H}$ and $F_{\rm HX}/F_{[\rm
O~III]}$ are significant. So we can accept the result.}. Their
obscuration seems to be close to that of face-on Sy1s. If the
scaleheight of the scattering zone varies with the central source
luminosity (Lumsden $\&$ Alexander 2001), the absence of their PBLs
may be due to either the small scaleheight in the scattering region
or the inexistence of BLRs. Since the less luminous non-HBLR Sy2s
have very small $L_{[\rm O~III]}$, their scattering screens may have
smaller scales than those of the luminous non-HBLR Sy2s or HBLR
Sy2s. However, the obscuration of this type of objects is very small
and seems to be the same as that of the host galaxy. So we suggest
that the invisibility of PBLs for less luminous non-HBLR Sy2s does
not arise from the obscuration.

The Eddington ratios of the less luminous non-HBLR Sy2s are
generally very small and their mean value is $10^{-2.57\pm0.27}$
(see Table 4). This is consistent with what Nicastro et al. (2003)
argued, that at very low accretion rates, the clouds of BLRs would
cease to exist. Since the obscuration of this type of objects is
very small, a key factor in the absence of PBLs is the very low
Eddington ratio rather than the obscuration. When the accretion rate
drops to extremely sub-Eddington values, their central engines
undergo fundamental changes and the BLR disappears (Ho 2008).
Recently, Tran et al. (2010) suggested that the low-luminosity AGNs
are probably powered by radiatively inefficient, or advection
dominated accretion flow (ADAF), that intrinsically lack BLRs, as
suggested observationally by e.g., Tran (2001, 2003); Bianchi et al.
(2008); Panessa et al. (2009); Shi et al. (2010), and inspired
theoretically by Nicastro (2000); Laor (2003); Elitzur $\&$ Shlosman
(2006); Elitzur $\&$ Ho (2009); Cao (2010).

In Table 4, we find that non-HBLR Sy2s can be classified into the
luminous ($L_{\rm [O~III]}>10^{41} \rm~ergs~s^{-1}$) and less
luminous samples, when considering only their obscuration. In light
of the above discussion, we hold that the invisibility of polarized
broad lines (PBLs) in the luminous non-HBLR Sy2s depends on the
obscuration; the invisibility of PBLs in the less luminous non-HBLR
Sy2s depends on the very low Eddington ratio rather than the
obscuration.

\section{Conclusion}

We conclude that HBLR Sy2s are dominated by AGNs, and non-HBLR Sy2s
are dominated by starbursts. This idea is supported by the evidences
listed below: (1) compared with non-HBLR Sy2s, HBLR Sy2s have larger
accretion rates and smaller $f_{60}/f_{25}$ ratio which may denotes
the relative strength of starbursts and AGN emissions; (2) HBLR Sy2s
are intrinsically more powerful than non-HBLR Sy2s from the analysis
of \Nev $\lambda 14.32$, \oiv $\lambda 25.89$, and \Neii $\lambda
12.81$, which are the useful single line diagnostics for
distinguishing AGN from starburst activity.

In addition, we find that the obscuration of less luminous non-HBLR
Sy2s is much smaller than that of luminous non-HBLR Sy2s or HBLR
Sy2s. We conclude that in luminous non-HBLR Sy2s, the invisibility
of PBLs is due to the obscuration (Shu et al. 2007); in less
luminous non-HBLR Sy2s, the invisibility of PBLs may not be due to
the scattering screen obscured by the obscuring material, but is
very likely due to the very low Eddington ratio and the BLRs are not
exist.

Although these results are from our large sample, we should further
consider sample completeness and have as large a sample size as
possible. In the future, both more complete and unbiased sample of
HBLR and non-HBLR Sy2s and fine measurements in various bands will
present the physical nature of non-HBLR and HBLR Sy2s.

\section*{ACKNOWLEDGMENTS}

referees for the careful
 reading of the manuscript and very helpful comments.
  We thank Chen Hu, and Xin-Lin Zhou for helpful suggestions
and discussions. We also thank James Wicker, Ali Tanni, Ping-Yan
Zhou, and Wei Du for polishing the English. This work was supported
by the Natural Science Foundation of China (NSFC) Foundation under
grants 10933001 and 10778726, the National Basic Research Program of
China (973 Program) No.2007CB815404, and the Young Researcher Grant
of National Astronomical Observatories, Chinese Academy of Sciences.

~~~~~~~~~~~~~~~~~~~~~~~~~~~~~~~~~~~~~~~~~~~~~~~~~~~~~~~~~~~~~~~~~~~~~~~~~~~~~~~~~~~~~~

~~~~~~~~~~~~~~~~~~~~~~~~~~~~~~~~~~~~~~~~~~~~~~~~~~~~~~~~~~~~~~~~~~~~~~~~~~~~~~~~~~~~~~

~~~~~~~~~~~~~~~~~~~~~~~~~~~~~~~~~~~~~~~~~~~~~~~~~~~~~~~~~~~~~~~~~~~~~~~~~~~~~~~~~~~~~~

~~~~~~~~~~~~~~~~~~~~~~~~~~~~~~~~~~~~~~~~~~~~~~~~~~~~~~~~~~~~~~~~~~~~~~~~~~~~~~~~~~~~~~

~~~~~~~~~~~~~~~~~~~~~~~~~~~~~~~~~~~~~~~~~~~~~~~~~~~~~~~~~~~~~~~~~~~~~~~~~~~~~~~~~~~~~~

~~~~~~~~~~~~~~~~~~~~~~~~~~~~~~~~~~~~~~~~~~~~~~~~~~~~~~~~~~~~~~~~~~~~~~~~~~~~~~~~~~~~~~

~~~~~~~~~~~~~~~~~~~~~~~~~~~~~~~~~~~~~~~~~~~~~~~~~~~~~~~~~~~~~~~~~~~~~~~~~~~~~~~~~~~~~~

~~~~~~~~~~~~~~~~~~~~~~~~~~~~~~~~~~~~~~~~~~~~~~~~~~~~~~~~~~~~~~~~~~~~~~~~~~~~~~~~~~~~~~

~~~~~~~~~~~~~~~~~~~~~~~~~~~~~~~~~~~~~~~~~~~~~~~~~~~~~~~~~~~~~~~~~~~~~~~~~~~~~~~~~~~~~~

~~~~~~~~~~~~~~~~~~~~~~~~~~~~~~~~~~~~~~~~~~~~~~~~~~~~~~~~~~~~~~~~~~~~~~~~~~~~~~~~~~~~~~

~~~~~~~~~~~~~~~~~~~~~~~~~~~~~~~~~~~~~~~~~~~~~~~~~~~~~~~~~~~~~~~~~~~~~~~~~~~~~~~~~~~~~~

~~~~~~~~~~~~~~~~~~~~~~~~~~~~~~~~~~~~~~~~~~~~~~~~~~~~~~~~~~~~~~~~~~~~~~~~~~~~~~~~~~~~~~

~~~~~~~~~~~~~~~~~~~~~~~~~~~~~~~~~~~~~~~~~~~~~~~~~~~~~~~~~~~~~~~~~~~~~~~~~~~~~~~~~~~~~~

~~~~~~~~~~~~~~~~~~~~~~~~~~~~~~~~~~~~~~~~~~~~~~~~~~~~~~~~~~~~~~~~~~~~~~~~~~~~~~~~~~~~~~

 \centerline{{\bf APPENDIX}}

\begin{table}[h] \caption{The non-HBLR Sy2 sample}

\renewcommand{\arraystretch}{1.2}

         \begin{tabular}{lccccccccl}
\hline

Name  & Reference$^{\rm a}$ & Name  & Reference$^{\rm a}$ & Name  &
Reference$^{\rm a}$ & Name  & Reference$^{\rm a}$ & Name
& Reference$^{\rm a}$ \\

(1) & (2) &(3) & (4) & (5) & (6)& (7) & (8)& (9) & (10) \\
\hline

ESO 428-G014& 3    & Mrk 334     & 24      & NGC1685     & 3      & NGC4501     & 5L      & NGC5695     & 3,5L  \\
F00198-7926 & 12   & Mrk 573     & 5L      & NGC2685     & 10     & NGC4565     & 10      & NGC5728     & 3,4A  \\
F01428-0404 & 10   & Mrk 938     & 5P      & NGC3031     & 10     & NGC4579     & 10      & NGC5929     &5P,6K,12\\
F03362-1642 & 5L   & Mrk 1066    & 2L      & NGC3079     & 5L     & NGC4594     & 10      & NGC6251     & 10    \\
F04103-2838 & 4A   & Mrk 1361    & 12      & NGC3147     &16KT,26K& NGC4698     & 16KT,26K& NGC6300     & 13A   \\
F04210+0401 & 4A   & NGC676      & 10      & NGC3281     & 3      & NGC4941     & 3       & NGC6890     & 3     \\
F04229-2528 & 4A   & NGC1058     & 10      & NGC3362     & 5L     & NGC5033     & 10      & NGC7130     & 12    \\
F04259-0440 & 12   & NGC1143     & 12      & NGC3393     & 17,11  & NGC5128     &  18A    & NGC7172     & 19A,12 \\
F08277-0242 & 4A   & NGC1144     & 5P      & NGC3486     & 10     & NGC5135     & 12,19A  & NGC7496     & 4A    \\
F10340+0609 & 3,8  & NGC1241     & 5P      & NGC3660     & 5L     & NGC5194     &   12    & NGC7582     & 19A,12 \\
F13452-4155 & 4A   & NGC1320     & 5L      & NGC3941     & 10     & NGC5256     &   12    & NGC7590     & 19A   \\
F19254-7245 & 14E  & NGC1358     & 3       & NGC3982     & 5L     & NGC5283     & 5L      & NGC7672     & 2L    \\
F20210+1121 & 4A   & NGC1386     & 3       & NGC4117     & 3      & NGC5347     & 5L      & NGC7679     & 10    \\
F23128-5919 & 4A   & NGC1667     & 3,5L    & NGC4472     & 10     & NGC5643     & 3       & UGC6100     & 5L    \\
IC 5298     & 12   & ...         &    ...  & ...         &    ... &  ...        & ...     & ...         &  ...  \\

 \hline  \multicolumn{10}{c}{The HBLR Sy2 sample}\\\hline
 \noalign{\smallskip}

Name  & Reference$^{\rm a}$ & Name  & Reference$^{\rm a}$ & Name  &
Reference$^{\rm a}$ & Name  & Reference$^{\rm a}$ & Name  &
Reference$^{\rm a}$ \\

(1) & (2) &(3) & (4) & (5) & (6)& (7) & (8)& (9) & (10) \\
\hline

ESO273-IG04  &   4A   & F18325-5926  &  13L   & MCG-3-58-7   & 5P     & NGC 591    &2L,3K  & NGC 5252   &  4A,15K\\
F00317-2142  &   10   & F20050-1117  &  10    & MCG-5-23-16  & 13A    & NGC 788    &  25L  & NGC 5506   &  5,13A \\
F00521-7054  &   4A   & F20460+1925  &  4A    & Mrk 3        & 2L     & NGC 1068   &  20L  & NGC 5995   &  12    \\
F01475-0740  &   5P   & F22017+0319  &  4A,5P & Mrk 78       & 2L     & NGC 2110   &  15K  & NGC 6552   &  5P    \\
F02581-1136  &   5L   & F23060+0505  &  7     & Mrk 348      & 2L     & NGC 2273   &  3K   & NGC 7212   &  1L    \\
F04385-0828  &   5LP  & IC 1631      &  10    & Mrk 463E     & 2L,4A  & NGC 2992   &  13A  & NGC 7314   &  13A   \\
F05189-2524  &   4A   & IC 3639      &  12    & Mrk 477      & 1L     & NGC 3081   &  3K   & NGC 7674   &  2L,4A \\
F11057-1131  &   4A   & IC 5063      & 13A,23A& Mrk 1210     & 1L     & NGC 3185   &  10   & NGC 7682   &  5P    \\
F15480-0344  &   4A   & Circinus     & 9E,21A & NGC 424      & 3C     & NGC 4388   &  4A   & Was 49b    &  1L    \\
F17345+1124  &   7    & MCG-3-34-64  &  4A    & NGC 513      & 22L    & NGC 4507   &  3K   & ...        &  ...   \\

            \hline
         \end{tabular}

\vskip 2mm

{ \noindent \vglue 0.3cm {\sc Notes}: Column 1, 3, 5, 7, and 9:
source name; Column 2, 4, 6, 8,
 and 10: the corresponding reference of the spectropolarimetric
 observations..

$^a$ Letters denote references that used the following telescope:
C=CTIO (4 m), P=Palomar (5 m), K=Keck (10 m), L=Lick (3 m), S=Subaru
(8.2 m), E=ESO (3.6), KT=Kitt (2.3 m), and A=AAT (3.9 m).

{\sc References}: (1) Tran 92; (2) Miller \& Goodrich 1990 ; (3)
Moran et al. 2000; (4) Young et al. 1996; (5) Tran 2001; (6) Moran
et al. 2001; (7) Gu \& Huang 2002; (8) Shu et al. 2007; (9) Oliva et
al. 1998; (10) Wang \& Zhang 2007; (11) Gu et al. 2001; (12) Lumsden
et al. 2001; (13) Lumsden et al. 2004; (14) Pernechele et al. 2003;
(15) Tran 2010; (16) Shi et al. 2010; (17) Nagao et al. 2000; (18)
Alexander et al. 1999; (19) Heisler et al. 1997; (20) Antonucci \&
Miller 1985; (21) Alexander et al. 2000; (22) Tran 1995; (23) Inglis
et al. 1993; (24) Ruiz et al. 1994; (25) Kay \& Moran 1998 (26) Tran
et al. 2010.}

\end{table}

\end{document}